\documentclass[aps,pra,twocolumn,groupedaddress]{revtex4-2}
\usepackage{graphicx,color}% Include figure files
\usepackage{amsmath}
\usepackage{amssymb}
\usepackage{bm}
\usepackage{braket}
\usepackage{here}
\usepackage{comment}

\begin{document}

%\preprint{}

\title{Kelvin wave in miscible two-component Bose-Einstein condensates}


\author{Kenichi Kasamatsu$^1$, Maki Okada$^1$, Hiromitsu Takeuchi$^2$}
%\thanks{}
\affiliation{$^1$Department of Physics, Kindai University, Higashi-Osaka, Osaka 577-8502, Japan \\
$^2$Department of Physics and Nambu Yoichiro Institute of Theoretical and Experimental Physics (NITEP), Osaka Metropolitan University, Osaka 558-8585, Japan}


\date{\today}

\begin{abstract}
We study the dispersion of Kelvin waves propagating along single- and half-quantum vortices in miscible two-component Bose-Einstein condensates based on the analysis of the Bogoloubov-de Genne equation. 
With the help of the interpolating formula connecting the dispersion relations at low- and high-wavenumber regime, we reveal the nontrivial dependence of the dispersion relation of the Kelvin waves on the intercomponent interaction through the change of the vortex core size of the vortical component. 
We also find the splitting of the Kelvin mode dispersion into gapless and gapfull branches when both components have overlapping single-quantized vortices. 
\end{abstract}

%\keywords{}

\maketitle

%\section{Introduction\label{intro}}

\section{Introduction}
Quantized vortices play a leading part of macroscopic quantum phenomena associated with superfluidity, thoroughly studied in, e.g, liquid heliums \cite{donnelly1991quantized}, cold atomic gases \cite{fetter2009rotating}, and exciton-polariton condensates \cite{lagoudakis2008quantized}. 
They are linear topological defects in three-dimensional superfluids and can possess axial wave excitations propagating along vortex lines. 
The gapless excitation involving helical deformation of a vortex line is known as the ``Kelvin wave" \cite{kelvin1880vibrations,pitaevskii1961vortex}. 
The Kelvin wave can exist in low-temperature superfluids, having been discussed intensively in the context of the decay mechanism of quantum turbulence \cite{kivotides2001kelvin,vinen2003kelvin,kozik2004kelvin,l2006energy,l2010spectrum,baggaley2011spectrum,sonin2012symmetry,autti2021vortex}. 
%Another gapless excitation involves the spatial oscillation of the core diameter along a vortex line, known as the ``varicose wave", having an analogy in the context of blood flow through vein \cite{wada1950standing}. 

A study of vortex waves in cold atomic Bose-Einstein condensates (BECs) has some advantages compared with that in the other superfluid systems. 
Experimentally, a vortex can be created in a well-controlled manner, e.g., by means of an external rotation \cite{madison2000vortex}, phase engineering \cite{matthews1999vortices}, and a moving obstacle \cite{neely2010observation}. 
Excitations of vortex waves are closely related with collective modes of the BEC \cite{bretin2003quadrupole}.
Also, vortex dynamics can be visualized by optical techniques, which enable us to make direct measurement of the three-dimensional dynamics of vortex lines \cite{serafini2015dynamics,serafini2017vortex}. 
Theoretically, the vortex states can be well described by the Gross-Pitaevskii (GP) mean-field theory \cite{fetter2001vortices}. 
The excitation spectrum of the Kelvin wave and the other vortex waves can be studied by the linearization analysis from the stationary solutions or direct numerical simulations of the GP equations \cite{fetter2004kelvin,simula2008kelvin,simula2008vortex,takeuchi2009spontaneous,simula2010kelvin,rooney2011suppression}. 

In this work, we discuss a nontrivial case, namely, Kelvin wave in miscible two-component BECs. 
Two-component (binary) BECs with tunable interatomic interactions have been realized in cold atomic gases \cite{papp2008tunable,tojo2010controlling,cabrera2018quantum,semeghini2018self}. 
The salient feature in this system is caused by the presence of the interatomic interactions between the different components, which determine, for example, the miscibility or immiscibility of the ground state structure. 
Although vortex dynamics in a binary superfluid system have been studied in various situations, most of them are restricted in the two-dimensional analysis \cite{skryabin2000instabilities,mcgee2001rotational,eto2011interaction,aioi2012penetration,ishino2013counter,kasamatsu2016short,wang2018dynamics,kuopanportti2019splitting,richaud2021dynamics,han2021annihilation,edmonds2021synthetic,ruban2021instabilities,ruban2021bubbles,han2022dynamics}. 
Although Hayashi \textit{et al}. considered, through the three-dimensional analysis, the dynamical instability of helical shear flows of binary BECs \cite{hayashi2013instability}, in which one flows along the core of the vortex line of the other component, their analysis was restricted to the immiscible case. 
 
A vortex in miscible two-component BECs takes a rich variety of core structures \cite{kasamatsu2005vortices}. 
When the first component has a vortex, the non-vortex second component is influenced by the presence of the vortex core; the density of the second component fills in the vortex core so that the superfluid order parameter does not vanish there.
As the second component feels the density distribution of the first component as the potential well, the second component forms a density peak (bottom) at the vortex axis for repulsive (attractive) intercomponent interaction. 
When there are vortices in both components, such an attractive interaction causes overlapping of the vortex cores \cite{kuopanportti2012exotic}. 
These variety of vortex structures make us expect the nontrivial dispersion relation of the vortex waves, compared with that of the single-component one. 
We calculate the dispersion relation of the Kelvin wave through the Bogoliubov--de-Genne (BdG) analysis to study the impact arising from intercomponent interaction to the properties of the Kelvin wave. 
We find that the Kelvin wave dispersion is well described by the formula that interpolates the quadratic$+$logarithmic form at a low-wavenumber regime and the quadratic form at a high-wavenumber regime, depending on the intercomponent interaction only through the vortex thickness. 

The paper is organized as follows. 
In Sec.~\ref{KWsingle}, we briefly review the properties of the Kelvin wave in a single-component BEC and introduce the interpolating method. 
Next, we turn to the problem of the two-component BECs, discussing dynamical stability of the vortex states in Sec.~\ref{Form} and detailed evaluation of the vortex core size in Sec.~\ref{Resultbdg}. 
The analyses of the BdG equations for the Kelvin wave are described in Sec.~\ref{honbanresult}. 
We conclude the paper in Sec.~\ref{concle}.

\section{Kelvin wave in a single-component Bose-EInstein condensate}\label{KWsingle}
First, we briefly review the Kelvin wave excitations in a single-component BEC based on the microscopic analysis of the BdG equation. 
We consider a straight vortex line in a dilute gaseous BEC confined in a cylinder. 
The vortex state can be obtained by solving the stationary GP equation $\left( \hat{h} + g |\Psi|^2 \right)\Psi = \mu \Psi$, where $\hat{h} = -\hbar^2 \nabla^2/(2M) + V_\text{ext}(\bm{r})$ is the single-particle Hamiltonian with the atomic mass $M$ and the external potential $V_\text{ext}(\bm{r})$. 
The parameter $g$ is the coupling constant, given by the $s$-wave scattering between cold atoms. 
The chemical potential $\mu$ determines the equilibrium bulk density $n = \mu/g$ far from the boundary. 
In the following, we consider the homogeneous system $(V_\text{ext}(\bm{r})=0)$ in the cylindrical coordinate $\bm{r} = (r,\theta,z)$ and impose the Neumann boundary condition at $r=R$.
%We assume that the external potential is given by the cylindrical box potential with the radius $R$ as
%\begin{align}
%V_\text{ext}(\bm{r}) = \biggl\{
%\begin{array}{cc}
%V_0 \quad & \text{for} \quad r > R \\
%0 \quad & \text{for}\quad  0 \leq r \leq R,
%\end{array}
%\label{cylinderpot}
%\end{align}
%where $V_0$ takes a sufficiently large value as $V_0=1000gn_0$ to represent the hard wall.

\begin{figure}[ht]
\centering
\includegraphics[width=0.9\linewidth]{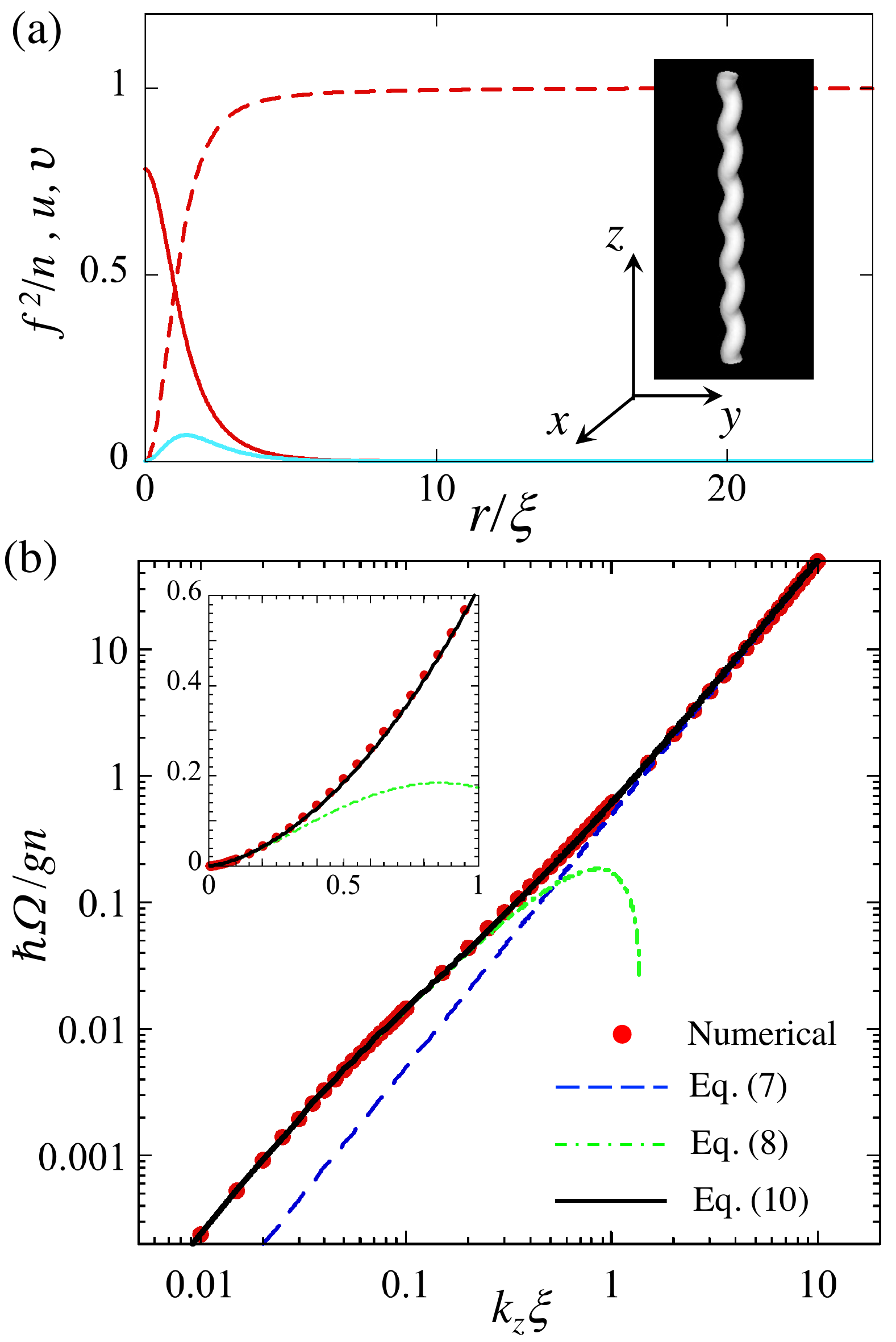} 
\caption{In (a), we plot the profile of the stationary solution with a vortex at $r=0$ (dashed curve) and the Bogoliubov amplitudes $u_{m l kz}(r)$ (dark red curve) and $v_{m l kz}(r)$ (light blue curve) with $(m,l,k_z)=(0,-1,\xi^{-1})$ for a single-component BEC in a cylinder; $u$ and $v$ are not normalized. 
The insets show the isosurface of $|\Psi+\delta \Psi|^2 = 0.8 n$ for the Kelvin mode with $k_z \xi = 1$ within the region $|z| \leq 20\xi$.
The panel (b) shows the dispersion relation of the Kelvin wave with $l=-1$ with $R=30\xi$. 
The result of the (red) dots is obtained by solving the BdG equation \eqref{BdGG1} numerically. 
We show the log-log plot of the data, where the small negative shift $\Delta/\mu \simeq - 10^{-3}$ at $k_z=0$ is subtracted. 
The (blue) dashed line represents $\hbar^2 k_z^2/(2M)$, while the (green) dashed-dotted curve represents Eq.~\eqref{Jelf} with the $r_\text{v}=0.7095\xi$. 
The (black) solid curve represents the interpolating formula of Eq.~\eqref{takeuchifit}. 
The inset shows the linear plot enlarged in the low-$k_z$ region.
}
\label{singler}
\end{figure}

The stationary vortex solution can be written as an axisymmetric form $\Psi (\bm{r}) = f(r) e^{i q \theta}$ with the real function $f(r)$ and the vortex winding number $q$; the solution has the translation symmetry along the $z$-axis. 
The radial profile of the vortex state with $q = 1$, depicted in Fig.~\ref{singler}(a), shows that the density becomes zero at the vortex core and heals from zero to the bulk value $n$ in the scale of the healing length $\xi = \hbar / \sqrt{M g n}$. 
We do not consider a vortex with $q\geq 2$, which is dynamically unstable \cite{pu1999coherent,mottonen2003splitting,kawaguchi2004splitting,shin2004dynamical,lundh2006dynamic,takeuchi2018doubly}. 

Next, we consider the fluctuation around the stationary vortex solution. 
The wave function is taken as $\Psi(\bm{r},t) = \Psi(\bm{r}) + \delta \Psi(\bm{r},t)$, where the fluctuation can be written as 
\begin{equation}
\delta \Psi (\bm{r},t) = e^{i q \theta - i \mu t/\hbar} \left[ u(\bm{r},t)-v^\ast (\bm{r},t) \right],
\end{equation}
and  the amplitudes $u(\bm{r},t)$ and $v(\bm{r},t)$ are set as
\begin{align}
u (\bm{r},t) =  \sum_{m,l,k_z} u_{mlk_z} (r) e^{i(l \theta + k_z z -\omega t)},   \label{bogo1u} \\
v (\bm{r},t) =  \sum_{m,l,k_z} v_{mlk_z} (r) e^{i(l \theta + k_z z -\omega t)} .\label{bogo1v}
\end{align}
Here, $m$ and $l$ represent the quantum number of the radial mode and the azimuthal mode, respectively, and $k_z$ the axial wave number along the $z$-axis. 
The radial profiles $u_{mlk_z} (r)$ and $v_{mlk_z} (r)$ depend on these quantum numbers, being obtained by solving the coupled BdG equation 
\begin{equation}
\mathcal{H}^{(lk_z)} \bm{w}_{mlk_z} = \hbar \omega \bm{w}_{mlk_z}.   \label{BdGG1}
\end{equation} 
This equation yields the eigenvalue $\omega = \omega_{mlk_z}$ and the eigenvector $\bm{w}_{mlk_z} = [u_{mlk_z}(r),v_{mlk_z}(r)]^\text{T}$.
The matrix $\mathcal{H}^{(lk_z)}$ is given by 
\begin{align}
\mathcal{H}^{(lk_z)}= \left(
\begin{array}{cc}
\hat{h}_{lk_z}^{+} & - g f(r)^2 \\ 
g f(r)^2 & -\hat{h}_{lk_z}^{-} 
\end{array}
\right)
\end{align}
with 
\begin{equation}
\hat{h}_{lk_z}^{\pm} =\frac{\hbar^2}{2M} \left( -\frac{\partial^2}{\partial r^2} - \frac{1}{r} \frac{\partial}{\partial r}+ \frac{(q \pm l)^2}{r^2} + k_z^2 \right) -\mu + 2 g f^2.
\end{equation} 

The normal Kelvin wave corresponds to the core-localized mode with the angular quantum number $l=-1$ for a vortex with $q=1$, whose dispersion relation and mode profile are shown in Fig.~\ref{singler}; these plots are obtained by solving numerically the BdG equation, the numerical details being addressed in the next section. 
The Bogoliubov amplitude is bounded at the vortex core, having a finite value at $r=0$, as shown in Fig.~\ref{singler}(a).
By taking account of the axial propagation $\propto e^{ik_z z}$ through Eqs.~\eqref{bogo1u} and \eqref{bogo1v}, the excitation actually involves helical deformation of the vortex core, as in the inset of Fig.~\ref{singler}(a). 
In Fig.~\ref{singler}(b), we plot the log-log plot of the dispersion relation. 
When the condensate is confined in the finite-size container, the dispersion relation of the Kelvin wave has a small negative shift $\Delta = - \hbar^2/(M R^{2})$ \cite{kobayashi2014kelvin,takahashi2015nambu}; in (b), we thus subtract $\Delta$ from $\hbar \omega$, i.e., $ \hbar \omega -\Delta  \equiv \hbar \Omega$, to make a log-log plot. 
The curve for $k_z \gg 1$ approaches asymptotically to the single-particle behavior 
\begin{equation}
\hbar \omega \simeq  \frac{\hbar^2 k_z^2}{2M}. \label{singledi}
\end{equation}
In the low-$k_z$ regime, the dispersion is given by the analytic expression derived in Ref.~\cite{takahashi2015nambu} as
\begin{equation}
\hbar \omega = \frac{\hbar^2 k_z^2}{2M} \left[ \ln \frac{1}{k_z r_\text{v}} - \chi(k_z R) \right].  \label{Jelf}
\end{equation}
Here, $r_\text{v}$ represents the cutoff length associated with the vortex core structure, whose value is given analytically as 
\begin{equation}
r_\text{v} = \frac{e^{0.577-0.227}}{2} \xi \simeq 0.7095 \xi. 
\end{equation}
In the following, we identify $r_\text{v}$ as a vortex core size. 
The function $\chi(k_z R)$ describes the correction of the finite-size effect, being given as $\chi(x) = [K_0(x) +K_2(x)]/[I_0(x)+I_2(x)]$ with the modified Bessel functions $I_n$ and $K_n$ of the first and second kind and having the asymptotic behavior $\chi(x) \sim 2/x^2$ for $x \ll1$ and $\chi(x) \sim \pi e^{-2x}$ for $x \gg 1$. 
Thus, the contribution $k_z^2 \chi(k_z R)$ takes a value $2/R^2$ at $k_z=0$ and converges rapidly to zero for $k_z \xi \gg \xi/R$ or $k_z R \gg 1$. 
%in the case of the harmonic trap potential used in the cold atom experiments,  see the discussion in Refs.~\cite{fetter2004kelvin,simula2008kelvin,simula2008vortex}. 
%Adapting $r_0 = 0.7095\xi$ following Ref.~\cite{takahashi2015nambu}, Eq.~\eqref{Jelf} describes well the numerical result for $k_z \xi \lesssim 0.3$. 

From the above-mentioned asymptotic behaviors, the Kelvin wave dispersion in the full range of $k_z$ can be described by interpolating the two relations Eqs.~\eqref{singledi} and \eqref{Jelf}. 
Among the types of simple interpolating functions, we see in Appendix~\ref{App1} that the arctan-type function gives the smallest difference from the numerical results. 
Thus, we suggest an interpolating formula of the Kelvin wave dispersion as
\begin{equation}
\hbar \omega_\text{int} = \frac{\hbar^2 k_z^2}{2M} \left\{ \ln \left[ \frac{e}{\frac{2}{\pi} \arctan (\frac{\pi}{2}e r_\text{v} k_z)} \right] -  \chi(k_z R) \right\}  . 
\label{takeuchifit}
\end{equation}
For $r_\text{v} k_z \ll 1$ the analytical formula of Eq.~\eqref{Jelf} is reproduced, while for $r_\text{v} k_z \gg 1$ the quadratic relation $\hbar \omega_\text{int}  \sim \hbar^2 k_z^2/(2M)$ is obtained. 
Figure~\ref{singler}(b) shows that this interpolating formula reproduces the numerical result quite well. 
The difference between the numerical results and Eq.~\eqref{takeuchifit} is within 5\% in the relevant range of $k_z$, as seen in Appendix~\ref{App1}. 
We will apply this interpolating function Eq.~\eqref{takeuchifit} to characterize the dispersion relation of the Kelvin wave in two-component BECs. 

%Also, there is another characteristic gapless mode propagating along the vortex line, known as the varicose wave \cite{donnelly1991quantized,takeuchi2009spontaneous}. 
%This wave is the lowest mode with $l=0$ and propagates by keeping the density profile axisymmetric. 
%As shown in Fig.~\ref{singler}(a), the dispersion relation of the varicose wave at the long-wavelength limit is connected with that of phonons, namely $\omega \propto k_z$. 
%For small $k_z$, the mode function has a node at $r=0$ and extends into the bulk region. 
%When this mode is excited, the bulk phonon involves the oscillation of the vortex core diameter propagating along the line. 
%We find that, as the axial wave number increases above $\xi^{-1}$, the mode profile is localized around the vortex core, as shown in Fig.~\ref{singler}(d) \cite{inpre}. 
%Thus, the varicose wave is also one of the characteristic modes localized at the vortex core.  

\section{Vortex core size in two-component Bose-Einstein condensates}\label{Form0}
Next, we seek the counterpart of the Kelvin wave of the axisymmetric vortex states in the miscible two-component BECs. 
Our focus is to clarify the effects of the intercomponent interaction to the dispersion relation and the Bogoliubov amplitudes of the Kelvin mode. 
After introducing the formulation, we classify dynamically stable regimes of the vortex states in Sec.~\ref{Form}.

We expect that the dispersion relation of the Kelvin wave in two-component BECs can be also characterized by the vortex core size, based on the prospects: 
(i) The dispersion relation in a short wavelength limit would also behave as a single-particle excitation. 
(ii) The dispersion relation in a long wavelength limit would be influenced only through the cutoff length or the vortex core size, since the microscopic feature such as the internal structure of the vortex core would not have a direct impact to the large-scale collective dynamics. 
In the following, we will verify this expectation by evaluating the vortex core size from the stationary solution in Sec.~\ref{Resultbdg} and extending the trial interpolating function Eq.~\eqref{takeuchifit} to the case of miscible binary condensates. 
%In order to discuss the core size dependence, it is necessary to use Eq.~\eqref{takeuchifit} instead of Eq.~\eqref{Jelf}, since Eq.~\eqref{Jelf} is valid only at a long-wavelength regime so that we face the ambiguity to determine the valid range of Eq.~\eqref{Jelf} which depends on the vortex core size. 

\subsection{Dynamically stable vortices}\label{Form}
The stationary state of the two-component BECs can be described by the coupled GP equations 
\begin{equation}
\mu_j \Psi_j = \left( \hat{h}_j + \sum_{j'=1,2} g_{jj'} |\Psi_{j'}|^2 \right) \Psi_j , \quad j=1,2.   \label{2compgpe}
\end{equation}
Here, the single-particle hamiltonian of the component $j$ is $\hat{h}_j = -\hbar^2 \nabla^2/(2 M_j)+ V_\text{ext}^j(\bm{r})$ and the coupling constants are $g_{jj'}$.  
Similar to Sec.~\ref{KWsingle}, the external potential $V_\text{ext}^j(\bm{r})$ is taken to be zero and impose the Neumann boundary condition at $r=R$.
In the homogeneous system, the mean field theory predicts that two components experience phase separation when $g_{12}/\sqrt{g_{11} g_{22}} > 1$ is satisfied \cite{ao1998binary}. 
On the other side, for $g_{12} / \sqrt{g_{11} g_{22}} < -1$, the condensates undergo a focusing collapse. 
In this work, we confine ourselves to the miscible regime $-1 < g_{12} /\sqrt{g_{11} g_{22}} < 1$, in which one can consider safely the dynamically stable vortex states. 
In the following, we fix the density at the bulk region as $|\Psi_1|^2 = |\Psi_2|^2 = n$ for simplicity and use the physical units determined by $n$.
We here put $M_{1} = M_{2} = M$, $g_{11} = g_{22} = g$, $\mu_1=\mu_2= g n (1+\gamma)$ with a new parameter $g_{12}/g \equiv \gamma$. 
To this end, we present our results by using the units independent of $\gamma$, namely 
\begin{equation}
\xi = \frac{\hbar}{\sqrt{M g n}}, \quad\quad \tau=\frac{\hbar}{g n}
\end{equation}
for length and time, respectively, similar to those in Fig.~\ref{singler} for a single-component BEC. 
These scales are useful to make clear the effect of intercomponent interaction $\gamma$ to the properties of the Kelvin mode. 
%The chemical potential is determined so as to satisfy $|\Psi_j|^2 = 1$ at the bulk region. 
%Throughout this work, the radius of the system is taken as $R=55 \xi$, which is much larger than the vortex core size for the relevant parameters. 

To consider a small fluctuation around the stationary solution, we write the wave function as $\Psi_j(\bm{r},t) = \Psi_j(\bm{r}) + \delta \Psi_j(\bm{r},t)$. 
Here the stationary solutions can be written as 
\begin{equation}
\Psi_j (r,\theta,z,t) = f_j(r) e^{i q_j \theta - i \mu_j t }
\end{equation}
with the real radial functions $f_1(r)$ and $f_2(r)$, and the vortex winding number $q_1$ and $q_2$. 
The fluctuation can be written as 
\begin{equation}
\delta \Psi_j (\bm{r},t) = e^{i q_j \theta -i \mu_j t} \left[ u_j(\bm{r},t)-v_j^\ast (\bm{r},t) \right].
\end{equation}
Along the similar line of the discussion in Sec.~\ref{KWsingle}, we consider the fluctuation by putting the ansatz as
$u_j (\bm{r},t) = \sum_{m,l,k_z} u_{mlk_z}^{(j)} (r) e^{i(l \theta + k_z z -\omega t)} $ and $v_j (\bm{r},t) = \sum_{m,l,k_z} v_{mlk_z}^{(j)} (r) e^{i(l \theta + k_z z -\omega t)} $.
The resulting BdG equation for the radial component reads 
\begin{equation}
\mathcal{H}^{(lk_z)} \bm{w}_{mlk_z}(r) = \hbar \omega \bm{w}_{mlk_z}(r), 
\label{eq:BdG}
\end{equation}
where 
\begin{equation}
\bm{w}_{mlk_z}(r) = \left[ u_{mlk_z}^{(1)} (r) , v_{mlk_z}^{(1)} (r), u_{mlk_z}^{(2)} (r), v_{mlk_z}^{(2)} (r) \right]^\text{T},
\end{equation}
\begin{align}
\mathcal{H}^{(lk_z)}
=
\left( 
\begin{array}{cccc}
\hat{h}_{lk_z,1}^+ & - g f_1 ^2 &g_{12} f_1 f_2 & - g_{12} f_1 f_2 \\
 g f_1^2 & -\hat{h}_{lk_z,1}^- & g_{12} f_1 f_2 & - g_{12} f_1 f_2  \\
g_{12} f_1 f_2  & -g_{12} f_1 f_2 & \hat{h}_{lk_z,2}^+ & - g f_2^2 \\
g_{12} f_1 f_2 & - g_{12} f_1 f_2  & g f_2^2 & -\hat{h}_{lk_z,2}^- \\
\end{array} 
\right),
\label{eq:BdGmatrix}
\end{align}
\begin{align}
h_{lk_z,j}^{\pm}= \frac{\hbar^2}{2M} \left(- \frac{\partial^2}{\partial r^2} - \frac{1}{r} \frac{\partial}{\partial r} + \frac{(q_j \pm l)^2}{r^2} + k_z^2 \right)   \nonumber \\
-\mu_j +2 g f_j^2 + g_{12} f_{j'}^2,  \quad\quad j' \neq j.
\end{align}
We solve this eigenvalue equation numerically to obtain the eigenvalue $\omega_{mlk_z}$ and the eigenfunctions $u_{mlk_z}^{(j)} (r)$ and $v_{mlk_z}^{(j)}  (r)$ for given $l$ and $k_z$. 
%Then, it is convenient to replace the radial functions as, e.g., $\psi_j = \phi_j/r$, in which we can use the Direchlet boundary condition such as $\phi_j(r=0) = 0$ and improve the accuracy of the results.

For later discussion, it is useful to remind the excitation spectrum of the density oscillation for a homogeneous system without vortices ($q_1=q_2=0$). 
In our simplified parameters, the excitation spectrum with respect to the wave number $k$ in a homogeneous system is \cite{pethick2008bose}
\begin{equation}
(\hbar \omega)^2_{\pm} = \frac{\hbar^2 k^2}{2M} \left[  \frac{\hbar^2 k^2}{2M} + 2 n (g \pm g_{12}) \right].   \label{phonodis}
\end{equation}
Here, we have two branches associated with plus (minus) sign, corresponding to the in-phase (out-of-phase) oscillation of the two-component densities. 
In the low-$k$ limit, we have the phonon dispersion $\omega \simeq \sqrt{g n (1 \pm \gamma)/M} \: k$, while we have the single-particle spectrum $\omega \simeq \hbar^2 k^2/(2M)$ in the high-$k$ limit. 
For $\gamma > 0$, the out-of-phase branch approaches to the quadratic relation $\omega_- \propto k^2$ as $\gamma \to 1$. 
Similarly, the in-phase branch becomes $\omega_+ \propto k^2$ as $\gamma \to -1$. 
These behaviors imply the instability associated with the phase separation at $\gamma=1$ and focusing collapse at $\gamma=-1$.

\begin{table}[ht]
\caption{\label{diagram} The diagram of the dynamical stability of the axisymmetric vortex states in the two-component BECs.}
\begin{tabular}{l|c|c}
\hline
  &$ \: -1< \gamma<0$ \: & \: $0<\gamma<1$ \:   \\
  \hline
(A) $q_1=1$, $q_2=1$ & stable & unstable \\ 
\hline
(B) $q_1=1$, $q_2=0$ & stable & stable \\ 
\hline
(C) $q_1=1$, $q_2=-1$ & unstable & unstable  \\
\hline
\end{tabular}
\end{table}
Without loss of generality, we can consider the three cases of the axisymmetric vortex states with $(q_1,q_2) = (1, 1)$, $(1, 0)$, and $(1, -1)$.  
In the particular parameter regimes, these vortex states have dynamical instability associated with the appearance of the imaginary frequency of the Bogoliubov excitations, as summarized in Table~\ref{diagram}. 
We confine ourselves to consider the Kelvin waves in the dynamically stable state.
%First of all, we do not consider the immiscible regime $\gamma > 1$ in this work, since the immiscible equilibrium solutions cannot be obtained through the imaginary time propagation of the GP equation with the fixed chemical potential. 
%However, for $(q_1,q_2) = (1, 0)$ we can expect the vortex with a massive core is realized \cite{richaud2021dynamics,richaud2020vortices}, which can be obtained by the imaginary time propagation with fixed particle numbers. 
%The BdG analysis in such a case needs a careful discussion, which will be reported elsewhere.  

\subsection{Vortex core size}\label{Resultbdg}
\begin{figure}[ht]
\centering
\includegraphics[width=1.0\linewidth]{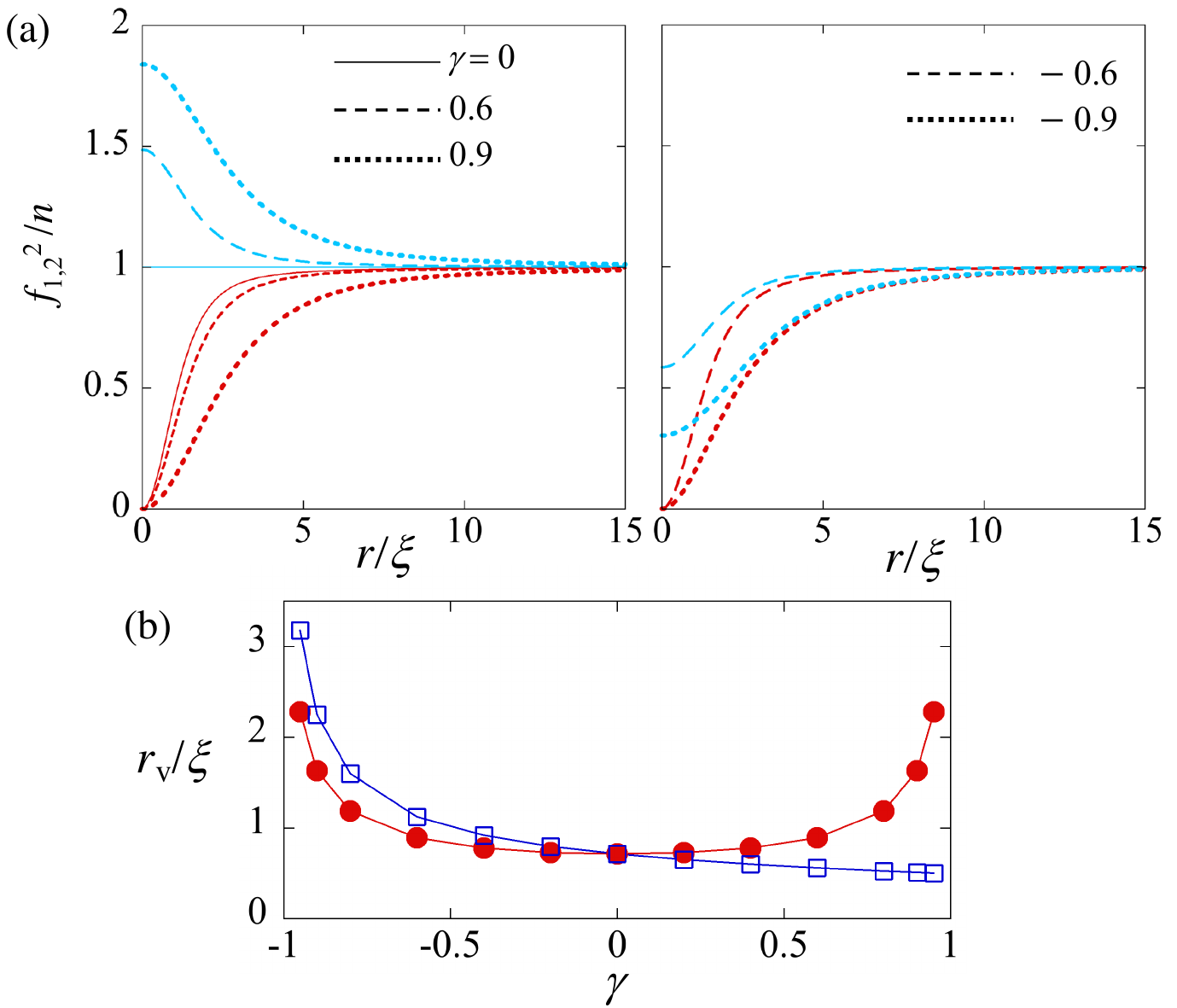} 
\caption{The panels (a) show the radial density profile of the axisymmetric vortex state with $(q_1,q_2) = (1,0)$ for several values of $\gamma$. 
The dark (red) and light (blue) curves represent $|\psi_1|^2$ and $|\psi_2|^2$, respectively and the left and right panels correspond to the solutions for $\gamma \geq 0$ and $\gamma <0$, respectively. 
The panel (b) shows the vortex core size, determined by the Gaussian fitting of the vortex core profile, for $(q_1,q_2) = (1,0)$ (filled circles) and $(q_1,q_2) = (1,1)$ (empty squares). 
The interpolating curves for the two plots are given by $r_\text{v} /\xi = 0.7095 / \sqrt{1-\gamma^2}$ and $r_\text{v}/\xi =  0.7095 /\sqrt{1+\gamma}$.
}
\label{stationaryfig}
\end{figure}
In the cases of (A) and (C) in Table~\ref{diagram}, the axisymmetric vortex states consist of singly-quantized vortices (SQV) with vanishing total densities at $r=0$, where the both components have exactly the same density profile $f_1(r)^2=f_2(r)^2$. 
Then, the nonlinear term of the GP equation can be written as $g(1+\gamma)f_j(r)^3$. 
As a result, the vortex core size is determined by the modified healing length $\xi/\sqrt{1+\gamma}$. 
To see this properties, we solve numerically Eq.~\eqref{2compgpe} and extract the vortex core size by making Gaussian fit $\propto e^{-r^2/(2\sigma^2)}$ to the profile $1-f_j^2/n$. 
For $\gamma=0$ we find that $\sigma \approx 1$ and the core size $r_\text{v}$ can be reproduced when multiplying $\sigma$ by a factor 0.7095. 
According to this fitting analysis and $r_\text{v} = 0.7095\sigma$, we determine the vortex core size for $\gamma \neq 0$, as shown in Fig.~\ref{stationaryfig}(b).
The core size for $\gamma>0$ $(\gamma < 0)$ decreases (increases) from that for $\gamma=0$, being written as 
\begin{equation}
r_\text{v} \simeq  \frac{0.7095 \xi}{\sqrt{1+\gamma}} \equiv r_\text{SQV} .
\end{equation} 
In other words, the vortex solutions for different values of $\gamma$ have profiles similar to Fig.~\ref{singler}(a) when we use the scaled coordinate $ r \sqrt{1+\gamma}/\xi$. 
For (A), the dynamically stable vortex configuration can only take place at $-1 < \gamma < 0$; for $0 < \gamma < 1$ the overlapping vortices experience the splitting dynamical instability \cite{kuopanportti2019splitting,han2022dynamics}. 
When $(q_1,q_2) = (1,-1)$, we have stationary profiles similar to the case (A), but the counter-rotating vortex state always gives rise to the dynamical instability for any values of $\gamma$ \cite{ishino2013counter,kuopanportti2019splitting,han2022dynamics}. 

For (B), we have a configuration of a half-quantized vortex (HQV) \cite{eto2011interaction,kasamatsu2016short} which is dynamically stable for $-1 < \gamma < 1$. 
Figure \ref{stationaryfig}(a) shows the stationary density profiles of the vortex states for $(q_1,q_2) = (1,0)$ for several values of $\gamma$. 
For $\gamma > 0$ the intercomponent interaction is repulsive so that the vortex core is filled by the other non-rotating component to reduce the overlapping of the condensate density. 
For $\gamma < 0$, on the other hands, the density of the non-rotating component is reduced together with the density depletion of the vortex core. 
The size of the vortex core of the $\Psi_1$-component has a nontrivial dependence of $\gamma$, as shown in Fig.~\ref{stationaryfig}(b), where we have done the Gaussian fitting analysis similar to the case (A) for the density profile $f_1^2$.
The core size behaves symmetrically with respect to sign of $\gamma$. 
This is due to two length scales in our problem, namely the `density' healing length $\xi_d = \xi / \sqrt{1+\gamma} $ and the `spin' healing length $\xi_s  = \xi / \sqrt{1-\gamma}$ \cite{eto2011interaction}. 
The former determines the spatial scale in which the total density $f_1^2 + f_2^2$ varies, and the latter does it for the density difference $f_1^2 - f_2^2$. 
Since the asymptotic behaviors of the profile functions for $r \gg 1$ are written as $(f_1^2+f_2^2)/n \sim 2-\xi_d^2/r^2$ and $(f_1^2 - f_2^2)/n \sim -\xi_s^2/r^2$ \cite{eto2011interaction}, the profile for 1st-component is written as $f_1^2/n \sim 1-(\xi_d^2+\xi_s^2)/(2r^2)$.
Thus, as shown in Fig.~\ref{stationaryfig}(b), the core size can be fitted as 
\begin{equation}
r_\text{v} = \frac{0.7095 \xi}{\sqrt{1-\gamma^2}} \equiv r_\text{HQV}.
\end{equation}

\section{Kelvin wave in two-component BECs}\label{honbanresult}
In this section, we discuss the properties of the Kelvin wave for each vortex state by solving the BdG equation \eqref{eq:BdG}. 
From the above discussions, the structure of the vortex core is relevant to properties of the Kelvin wave in two-component BECs through the ratio of the core size $r_\text{v}=r_\text{v}(\gamma)$ and the system size $R$ as a finite size effect. 
We thus fix the system size for a certain $\gamma$ to be $R/\xi = 30 r_\text{v}(\gamma)/r_\text{v}(0)$ throughout the following discussion to compare the results of the Kelvin wave for $\gamma=0$.

\subsection{Kelvin wave of a HQV}\label{Resultbdg10}
\begin{figure}[ht]
\centering
\includegraphics[width=0.84\linewidth]{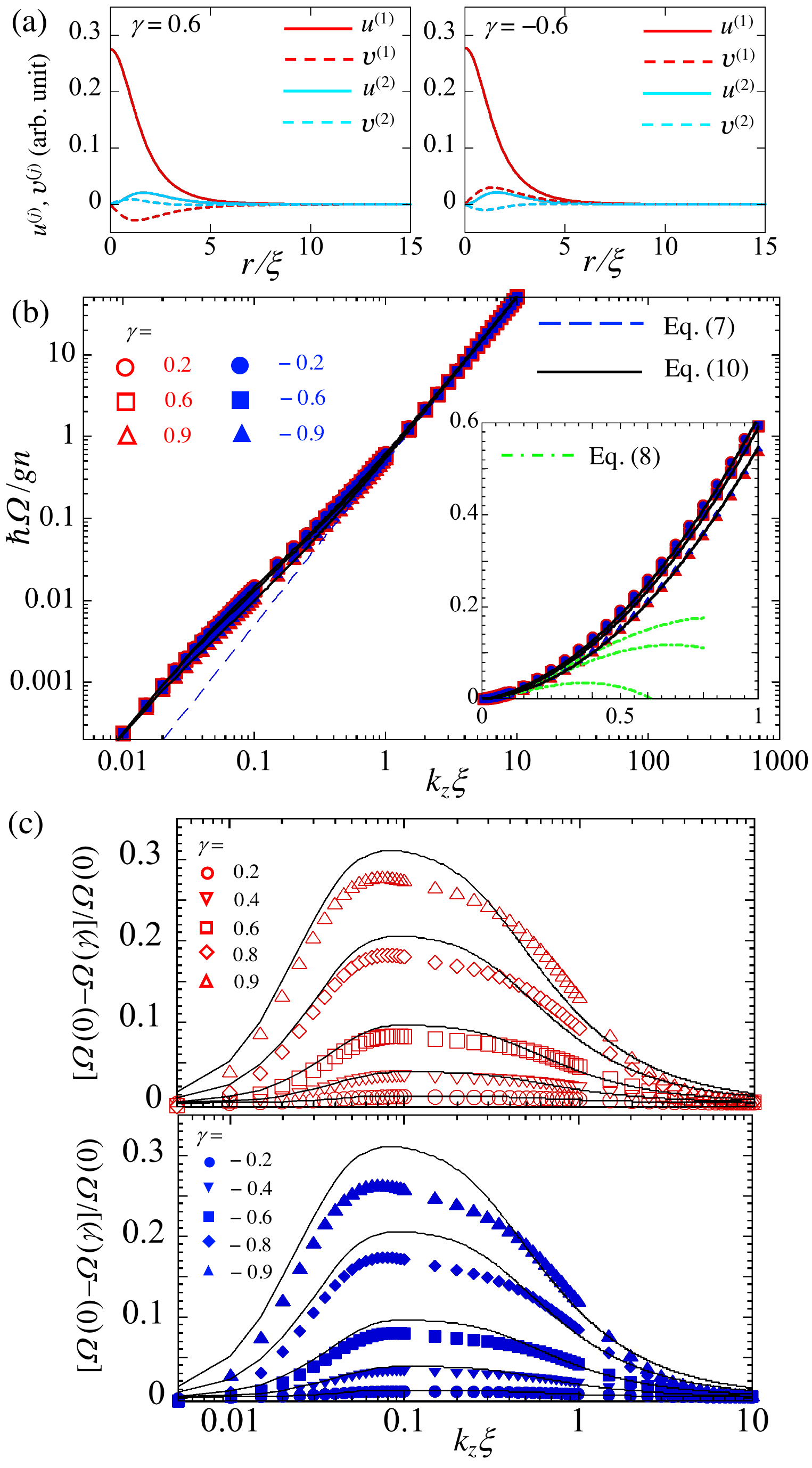} 
\caption{The properties of the Kelvin wave ($l=-1$) for $(q_1,q_2)=(1,0)$. 
(a) The Bogoliubov amplitudes $u_{m l kz}^{(1)}(r)$ (dark red curve), $v_{m l kz}^{1} (r)$ (light blue curve), $u_{m l kz}^{(2)}(r)$ (dark red dashed curve), $v_{m l kz}^{2} (r)$ (light blue dashed curve) with $(m,l,k_z)=(0,-1,\xi^{-1})$, which are typical for $\gamma>0$ (left) and $\gamma<0$ (right).
(b) Log-log plots of the Kelvin wave dispersion are for several values of $\gamma$. 
The red open symbols for $\gamma>0$ are almost coincident with the blue filled sybmbols for $\gamma<0$ with the same magnitude $|\gamma|$. 
The solid, dashed, and dashed-dotted curves refer to Eqs.~\eqref{takeuchifit}, \eqref{singledi}, and \eqref{Jelf}, respectively. 
The inset shows the linear plot enlarged in the low-$k_z$ region, where the green dashed-dotted curves are Eq.~\eqref{Jelf} with $r_\text{v} = r_\text{HQV}$ for $|\gamma| = 0.2$, 0.6, and 0.9 from the top to bottom curves. 
In (c), we plot the difference of the dispersion relation from that of $\gamma=0$. 
The solid curves represent the corresponding difference calculated from the interpolating formula Eq.~\eqref{takeuchifit} with $r_\text{v}=r_\text{HQV}$. 
The upper and lower panels show the results for $\gamma > 0$ and $\gamma < 0$, respectively.
 } 
\label{disp1}
\end{figure}
First, we consider the Kelvin wave for $(q_1,q_2)=(1,0)$, i.e., a half-quantized vortex. 
As in the single-component BEC, the Kelvin mode corresponds to the lowest-energy mode with $l=-1$. 
As shown in Fig.~\ref{disp1}(a), the Bogoliubov amplitude, especially $u^1_{mlk_z}$, of the Kelvin mode is localized at the vortex core. 
For $\gamma>0$ the amplitude $u^2_{mlk_z}$ of the non-vortex component takes the opposite sign of $u^1_{mlk_z}$  [the left panel of Fig.~\ref{disp1}(a)]. 
This means that the density hump of $\Psi_2$ filling the vortex core of $\Psi_1$ catches up with the displacement of the vortex core when the Kelvin wave is excited. 
For $\gamma < 0$ the Bogoliubov amplitudes distribute similarly to the case of $\gamma>0$, but the amplitude $u^2_{mlk_z}$ takes the same sign of $u^1_{mlk_z}$ [the right panel of Fig.~\ref{disp1}(a)]. 
This also insists that the density hollow of $\Psi_2$ at the vortex core of $\Psi_1$ catches up with the displacement of the vortex core when the Kelvin wave is excited. 

Figure \ref{disp1}(b) shows the dispersion relation of the Kelvin wave for $(q_1,q_2) = (1, 0)$ and various values of $\gamma$. 
To make a log-log plot, we subtract the negative energy shift $\Delta$ at $k_z = 0$ as $\hbar \omega(\gamma)-\Delta(\gamma) = \hbar \Omega(\gamma)$ as in Fig.~\ref{singler}(b).
All plots at $k_z \xi \gg 1$ approach asymptotically to the quadratic function $\hbar^2 k_z^2/(2M)$, independent of the values of $\gamma$. 
This is because for $k_z \xi \gg 1$ the term $\hbar^2 k_z^2/(2M)$ in the diagonal component of Eq.~\eqref{eq:BdGmatrix} becomes dominant. 
%Since the tension of the vortex (an energy of a vortex per unit length) with $(q_1,q_2) = (1, 0)$ is written as $(\pi \hbar^2 n/ M) \ln R/\xi $ \cite{eto2011interaction}, the excitation energy of the Kelvin wave is roughly estimated by the replacement $R \sim 1/k_z$ \cite{donnelly1991quantized}. 
%Because of our scaled unit based on the fixed bulk density $n$, the prefactor of the dispersion relation is independent of $\gamma$.
%Since the tension of the vortex (an energy of a vortex per unit length) with $(q_1,q_2) = (1, 0)$ is written as $(\pi \hbar^2 n/ M) \ln R/\xi $ \cite{eto2011interaction}, the excitation frequency of the Kelvin wave with the wavelength $\lambda$ is roughly determined by its factor as $(\pi \hbar^2 n/ M) \times \lambda \sim \hbar^2 k_z^2/M$ (under the assumption $n \lambda^3 \sim \mathcal{O}(1)$), being insensitive to the change of $\gamma$ due to our scaled unit based on the fixed bulk density $n$. 
However, the curvature of the dispersion curve at the low-$k_z$ region has a weak dependence on $\gamma$, as seen in the inset of Fig.~\ref{disp1}(b). 
We find that the analytic expression Eq.~\eqref{Jelf}, in which $r_\text{v}$ is replaced by $r_\text{HQV}$ instead of $0.7095\xi$, can describe well the dispersion at a low-$k_z$ regime.
To show this result more clearly, we plot in Fig.~\ref{disp1}(c) the difference of the dispersion relation from that for $\gamma=0$, i.e., that of the single-component BEC. 
Here, the difference is evaluated by $[\Omega(0)-\Omega(\gamma)]/\Omega(0)$. 
These $\gamma$ dependence can be explained by the fact that the vortex core size with $(q_1,q_2) = (1, 0)$ increases together with $\gamma$. 
To confirm this, we plot the similar plot, but now it is calculated from the interpolating formula of Eq.~\eqref{takeuchifit}. 
Here, the core size $r_\text{v}$ in Eq.~\eqref{takeuchifit} is also evaluated by $r_\text{v}=r_\text{HQV}$. 
The curves in Fig.~\ref{disp1}(b) can capture the obtained $\gamma$ dependence quite well. 

We also note that the dispersion relations for $\gamma<0$ are almost identical to those for $\gamma>0$ with the same magnitude, as seen in Fig.~\ref{disp1}(b) and (c). 
Thus, the dispersion relation behaves symmetrically with respect to the sign of $\gamma$, which supports to conclude that the dispersion relation of the Kelvin wave is characterized by the vortex core size of the vortical component, as expected from the $\gamma^2$-dependence of $r_\text{v}$ in Fig.~\ref{stationaryfig}(b). 

%The right inset shows the mode amplitude $(u_1,v_1)$, which is localized around the vortex core. 
%The width increases with $\gamma$, since the size of the vortex core increases. 
%We note that the second lowest mode with $l=-1$ tends to behave similarly with the Kelvin mode as $\gamma \to 1$ due to the strong coupling between two components. 

%The dispersion relation of the varicose mode [Fig.~\ref{disp1}(b)] at low-$k$ region can be described by the out-of-phase phonon dispersion of Eq.~\eqref{phonodis}; this axial compressional mode is irrelevant to the existence of a vortex. 
%As $\gamma$ approaches to unity, the dispersion at low $k$ region behaves from linear to quadratic. 
%The localization behavior can be seen from the profile of the Bogoliubov amplitude $(u_1,v_1)$. 
%The amplitude at $k=1$ tends to delocalize with increasing $\gamma$, approaching the bulk phonon excitation. 
%This is because with increasing $\gamma$ the excitation tends to be hybridized strongly with those of the other non-vortex component, which are always delocalized. 
%We find that the localization behavior takes place for higher $k_z$ with increasing $\gamma$. 
%Note that, despite the localization of mode function around the core, the bulk phonon mode also induces the modulation of the vortex core diameter along the vortex axis. 

\subsection{Kelvin wave of a SQV}\label{Resultbdg11}
Next, we consider the Kelvin wave for $(q_1,q_2) = (1, 1)$.
In this setting, the vortex state is dynamically stable only for  $\gamma \leq 0$.  
Since both components have a vortex, we can consider two branches of the Kelvin wave with node-less radial modes, corresponding to the in-phase and out-of-phase oscillation (with $\pi$-phase difference) of the helical vortex-line deformation. 
For $\gamma < 0$, the intercomponent attractive interaction favors the in-phase oscillation rather than the out-of-phase one, which is generally gapfull since it involves the deformation of the vortex core profile. 
We here discuss the two branches, since the both branches are gapless for $\gamma=0$ and the gapfull mode could be alive as a low-energy excitation when $\gamma$ is sufficiently small. 

%We confirm that the Bogoliubov amplitudes take $(u_1,v_1) = (u_2,v_2)$ for the in-phase mode and $(u_1,v_1) = - (u_2,v_2)$ for the out-of-phase mode, although we do not explicitly in the following.

\begin{figure}[ht]
\centering
\includegraphics[width=1.0\linewidth]{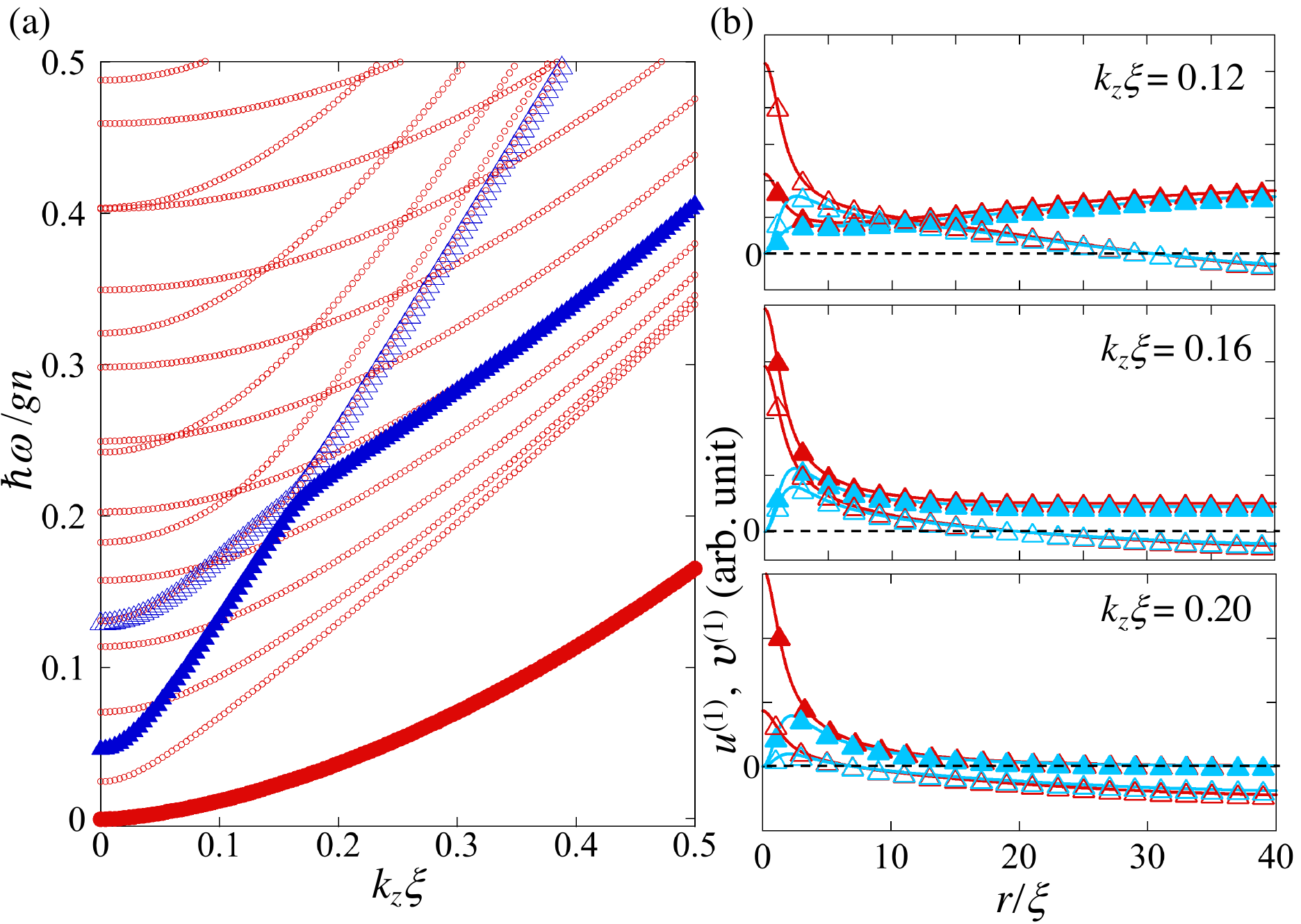} 
\caption{The panel (a) shows the dispersion relation of the Bogoliubov modes with $l=-1$ and various values of the radial quantum number $m$ for $(q_1,q_2) = (1,1)$ and $\gamma=-0.6$. 
The lowest mode (filled red circles) represents the in-phase Kelvin mode. 
The out-of-phase Kelvin mode emerges at $k_z \xi \gtrsim 0.16$ as a result of the avoided crossing of the two out-of-phase collective modes, shown by the blue filled and empty triangles.
In (b), we show the radial profile of the Bogoliubov modes $(u^{(1)}_{mlk_z},v^{(1)}_{mlk_z})$ relevant to the avoided crossing in (a) for $k_z \xi= 0.12$, 0.16, and 0.2. 
The dark-red curve and the light-blue one with filled triangles correspond to the the amplitudes $u^{(1)}_{mlk_z}$ and $v^{(1)}_{mlk_z}$, respectively, of the lower-lying mode [filled triangles in (a)], while those with empty triangles correspond to the higher-lying mode [empty triangles in (a)].
The profile of the second component is out-of-phase $(u^{(2)}_{mlk_z},v^{(2)}_{mlk_z})=-(u^{(1)}_{mlk_z},v^{(1)}_{mlk_z})$, not shown here. 
}
\label{avoidcrossss}
\end{figure}
Figure \ref{avoidcrossss}(a) shows a series of eigenvalues of the Bogoliubov modes with $l=-1$ for $\gamma=-0.6$ as a function of $k_z \xi$. 
The in-phase Kelvin mode corresponds to the lowest gapless mode, which is well separated from the other gapped excitation branches. 
This dispersion relation is well described by the interpolating formula of Eq.~\eqref{takeuchifit} with a suitable choice of $r_\text{v}$ (see the following discussion).
The out-of-phase Kelvin mode appears above a certain axial wavenumber $k_z \xi$ as a result of an avoided crossing of the two out-of-phase collective modes which are extended over the system in the low-$k_z$ limit. 
As shown in Fig.~\ref{avoidcrossss}(a), the avoided crossing takes place at $k_z \xi= 0.16$, below which the two relevant modes are extended to the bulk region [top panel of Fig.~\ref{avoidcrossss}(b)]. 
For $k_z  \xi>0.16$ the dispersion curve of the lower-lying mode approaches to the quadratic form and concurrently its mode amplitudes are localized at the vortex core, as seen in the bottom panel of Fig.~\ref{avoidcrossss}(b). 

\begin{figure}[ht]
\centering
\includegraphics[width=0.8\linewidth]{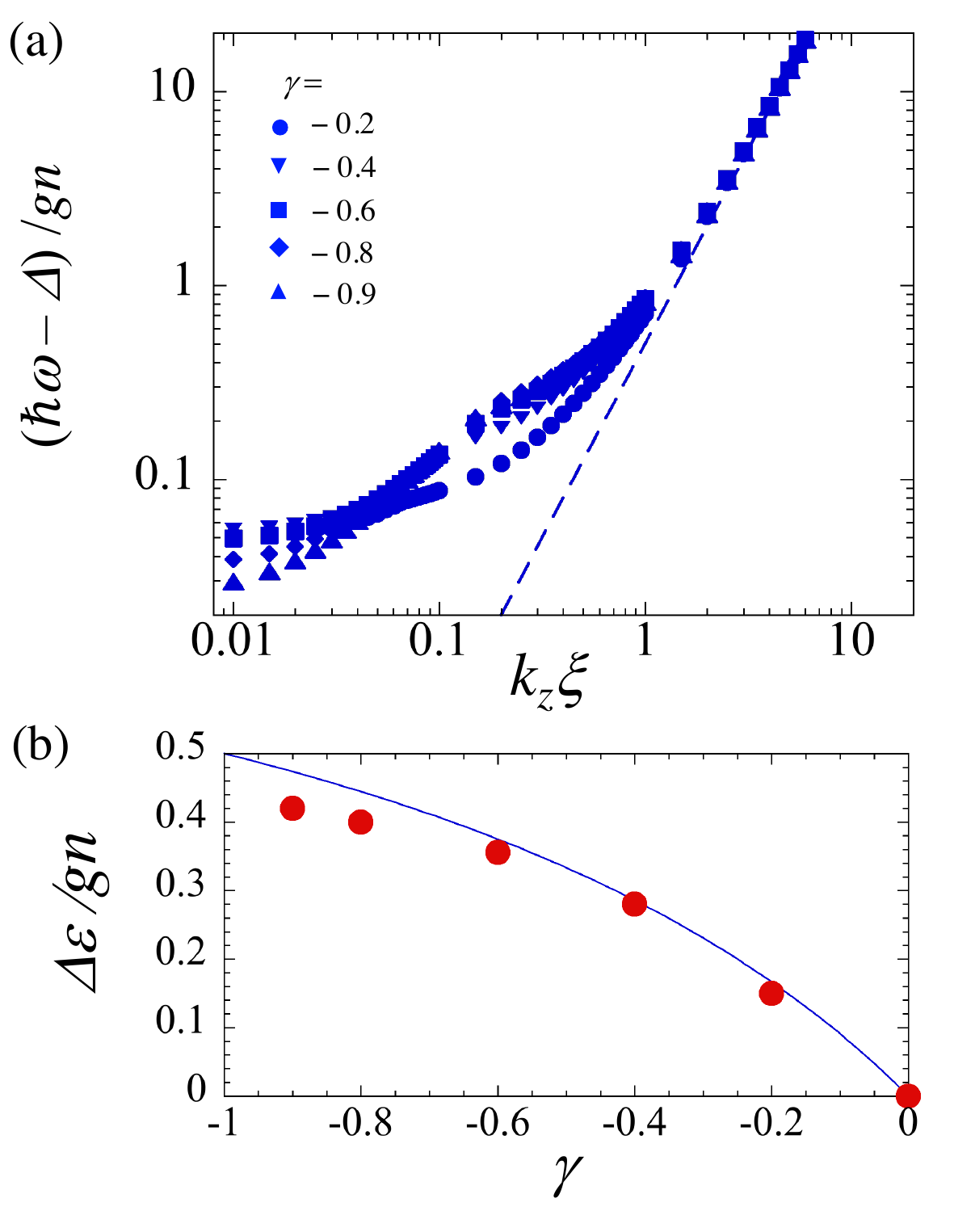} 
\caption{In (a), log-log plots of the dispersion relations of the out-of-phase Kelvin mode are shown for several values of $\gamma (<0)$. 
The (blue) dashed line represents $\hbar^2 k_z^2/(2M)$. 
The curves in the lower-$k_z$ region represent the frequency of the lower-lying out-of-phase mode contributing the avoided crossing. 
The panel (b) shows the energy gap of the out-of-phase mode, extracted by the relation $\hbar \omega = \hbar^2 k_z^2/(2M) + \Delta \epsilon$ at the high-$k_z$ region. 
The solid curve represents $ \Delta \epsilon/(gn) = - \gamma/(1-\gamma)$.
}
\label{doublevorcase}
\end{figure}
%Figure \ref{doublevorcase} (a) shows the dispersion relations of the in-phase Kelvin mode for several values of $\gamma (< 0)$. 
%The in-phase mode has properties similar to the case of $(q_1,q_2)=(1,0)$ in Sec.~\ref{Resultbdg10} with respect to $\gamma$. 
%The dispersion curve has not a significant dependence on $\gamma$, approaching to the quadratic form for $k_z \xi \gg 1$. 
%Applying the interpolating formula of Eq.~\eqref{takeuchifit} with $r_\text{v} = r_\text{SQV}$, we have a result similar to Fig.~\ref{disp1}(c) (not shown). 
%Thus, the Kelvin wave dispersion at a low-$k_z$ regime is modified by $\gamma$ through the vortex core size. 
%

The dispersion relation of the in-phase mode for $\gamma < 0$ takes the form similar to that of the single-component BEC, being just written by Eq.~\eqref{takeuchifit} with $r_\text{v} = r_\text{SQV}$. 
This is because Eq.~\eqref{eq:BdG} with the conditions $f_1=f_2$ and $(u^1_{mlk_z},v^1_{mlk_z})=(u^2_{mlk_z},v^2_{mlk_z})$ owing to the in-phase mode reduce to the single-component BdG equation, so that one can reproduce the unique dispersion curve seen in Fig.~\ref{singler} under a suitable scaling plot, apart from the finite size correction $\chi(k_z R)$. 
Figure~\ref{doublevorcase} (a) shows the dispersion relations of the out-of-phase Kelvin mode for several values of $\gamma$, which also exhibit the quadratic dependence with respect to $k_z$ at $k_z \xi \gg 1$. 
Thus, the out-of-phase mode at high-$k_z$ region can be written as $\hbar \omega \simeq \hbar^2 k_z^2/(2M) + \Delta \epsilon$ with a constant energy gap $\Delta \epsilon$. 
The energy gap arises from the fact that the relative displacement of the vortex cores in each component involves the deformation of the core structure and results in the energetic cost. 
In (b), we plot the energy gap $\Delta \epsilon$ as a function of $\gamma$. 
This $\gamma$-dependence of $\Delta \epsilon$ can be fitted well by the relation $\Delta \epsilon \propto - \gamma/(1-\gamma)$, whose derivation needs more detailed consideration about short-range properites of the vortex-vortex interaction.

\section{Conclusion}\label{concle}
In summary, we discussed the Kelvin wave of SQV and HQV in miscible two-component BECs. 
We first confirmed that the Kelvin wave dispersion of a single-component BEC is well described by the interpolating formula Eq.~\eqref{takeuchifit} in a whole range of $k_z$. 
Based on this interpolating formula and the precise evaluation of vortex core properties, we considered the impact of the intercomponent interaction on the Kelvin mode by solving the BdG equation. 
For $(q_1,q_2)=(1,0)$, the Kelvin wave dispersion is weakly dependent on the intercomponent interaction only through the change of the vortex core size in the vortical component, being written by Eq.~\eqref{takeuchifit} with $r_\text{v} = r_\text{HQV}$. 
Thus, the dispersion is symmetric with respect to the sign of the intercomponent coupling constant. 
In the case of $(q_1,q_2)=(1,1)$ and the attractive intercomponent interaction $\gamma <0$, we have both lower-lying in-phase and higher-lying out-of-phase branches for the Kelvin wave. 
The dispersion of the in-phase branch is gapless, being also written by Eq.~\eqref{takeuchifit} with $r_\text{v} = r_\text{SQV}$.
The out-of-phase Kelvin wave is gapfull excitation, being generated from the avoided crossing of the two out-of-phase delocalized modes as low-$k$. 
This energy gap is associated with the deformation of the vortex core caused by the relative displacement of the vortex position from the center. 

\begin{acknowledgments}
This research was supported by JSPS KAKENHI Grants No. JP18KK0391, No. JP20H01842, and No. 20H01843 and in part by the 2022 Osaka Metropolitan University (OMU) Strategic Research Promotion Project (Priority Research). 
%The work of K.K. is supported by KAKENHI from the Japan Society for the Promotion of Science (JSPS) Grant-in- Aid for Scientific Research (KAKENHI Grant No. 18K03472).
\end{acknowledgments}

\appendix
\section{How to determine an interpolating function between Eqs.~\eqref{singledi} and \eqref{Jelf}} \label{App1}

\begin{figure}[ht]
\centering
\includegraphics[width=0.8\linewidth]{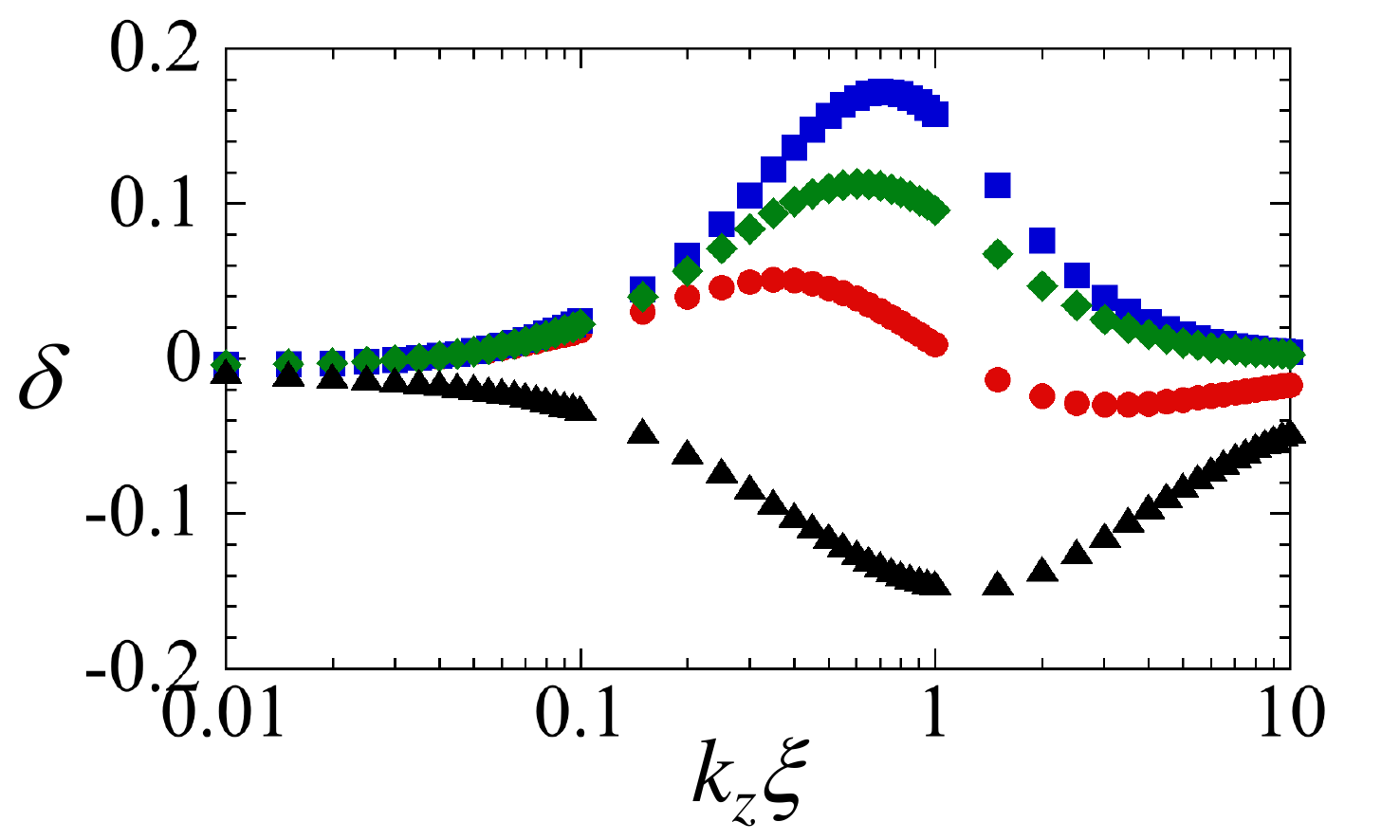} 
\caption{
The figure shows the errors between the numerical data and interpolating formula given by (A) $\arctan{x}$ (red-curcles), (B) $\tanh x$ (blue-squares), (C) $x/\sqrt{1+x^2}$ (green-diamonds), and (D) $x/(1+x)$ (black-triangles). 
Here, the error is defined by $\delta \equiv (\Omega - \Omega_\text{int})/\Omega$.
}
\label{singler2a}
\end{figure}
We here describe how to determine the interpolating functions between Eqs.~\eqref{singledi} and \eqref{Jelf} among several possible interpolating function. 
To interpolate Eqs.~\eqref{singledi} and \eqref{Jelf}, the logarithmic term of the dispersion relation of the Kelvin wave can be modified as 
\begin{equation}
\hbar \omega_\text{int} = \frac{\hbar^2 k_z^2}{2M} \left[ \ln \frac{e}{\phi(e k_z r_\text{v})} -\chi(k_z R) \right].
\end{equation}
Here, $\phi(x)$ $(x>0)$ is an interpolating function satisfying the asymptotic behavior $\phi(x) \sim x$ for $x \ll 1$ and $\phi(x) \sim 1$ for $x \gg 1$. 
Note that the contribution $\chi(k_z R)$ from the finite-size effect converges to zero for $k_z R \gg 1$, so that we need not to consider it in the asymptotic limit at $x \gg 1$.
We consider several simple functions satisfying the above asymptotic behavior: (A) $\phi(x) = (2/\pi) \arctan(\pi x/2)$, (B) $\phi(x) = \tanh(x)$, (C) $\phi(x) = x/\sqrt{1+x^2}$, (D) $\phi(x) = x/(1+x)$. 
To seek the better choice that can reproduce the numerical results, we calculate the deviation of the interpolating function from the numerical one as $\delta = (\Omega - \Omega_\text{int})/\Omega$, where $\hbar \Omega = \hbar \omega - \Delta$ with the negative energy shift $\Delta$ at $k_z=0$ (see the text). 
Among the above lists, the choice (A) gives a better interpolation within the 5\% error in the intermediate range of $k_z \xi$, as shown in Fig.~\ref{singler2a}. 
We thus adopt Eq.~\eqref{takeuchifit} in the analysis.

%\begin{acknowledgments}
%\end{acknowledgments}

\bibliography{reference}

\providecommand{\noopsort}[1]{}\providecommand{\singleletter}[1]{#1}%
\begin{thebibliography}{57}%
\makeatletter
\providecommand \@ifxundefined [1]{%
 \@ifx{#1\undefined}
}%
\providecommand \@ifnum [1]{%
 \ifnum #1\expandafter \@firstoftwo
 \else \expandafter \@secondoftwo
 \fi
}%
\providecommand \@ifx [1]{%
 \ifx #1\expandafter \@firstoftwo
 \else \expandafter \@secondoftwo
 \fi
}%
\providecommand \natexlab [1]{#1}%
\providecommand \enquote  [1]{``#1''}%
\providecommand \bibnamefont  [1]{#1}%
\providecommand \bibfnamefont [1]{#1}%
\providecommand \citenamefont [1]{#1}%
\providecommand \href@noop [0]{\@secondoftwo}%
\providecommand \href [0]{\begingroup \@sanitize@url \@href}%
\providecommand \@href[1]{\@@startlink{#1}\@@href}%
\providecommand \@@href[1]{\endgroup#1\@@endlink}%
\providecommand \@sanitize@url [0]{\catcode `\\12\catcode `\$12\catcode
  `\&12\catcode `\#12\catcode `\^12\catcode `\_12\catcode `\%12\relax}%
\providecommand \@@startlink[1]{}%
\providecommand \@@endlink[0]{}%
\providecommand \url  [0]{\begingroup\@sanitize@url \@url }%
\providecommand \@url [1]{\endgroup\@href {#1}{\urlprefix }}%
\providecommand \urlprefix  [0]{URL }%
\providecommand \Eprint [0]{\href }%
\providecommand \doibase [0]{https://doi.org/}%
\providecommand \selectlanguage [0]{\@gobble}%
\providecommand \bibinfo  [0]{\@secondoftwo}%
\providecommand \bibfield  [0]{\@secondoftwo}%
\providecommand \translation [1]{[#1]}%
\providecommand \BibitemOpen [0]{}%
\providecommand \bibitemStop [0]{}%
\providecommand \bibitemNoStop [0]{.\EOS\space}%
\providecommand \EOS [0]{\spacefactor3000\relax}%
\providecommand \BibitemShut  [1]{\csname bibitem#1\endcsname}%
\let\auto@bib@innerbib\@empty
%</preamble>
\bibitem [{\citenamefont {Donnelly}(1991)}]{donnelly1991quantized}%
  \BibitemOpen
  \bibfield  {author} {\bibinfo {author} {\bibfnamefont {R.~J.}\ \bibnamefont
  {Donnelly}},\ }\href@noop {} {\emph {\bibinfo {title} {Quantized vortices in
  helium II}}},\ Vol.~\bibinfo {volume} {2}\ (\bibinfo  {publisher} {Cambridge
  University Press},\ \bibinfo {year} {1991})\BibitemShut {NoStop}%
\bibitem [{\citenamefont {Fetter}(2009)}]{fetter2009rotating}%
  \BibitemOpen
  \bibfield  {author} {\bibinfo {author} {\bibfnamefont {A.~L.}\ \bibnamefont
  {Fetter}},\ }\bibfield  {title} {\bibinfo {title} {Rotating trapped
  bose-einstein condensates},\ }\href@noop {} {\bibfield  {journal} {\bibinfo
  {journal} {Reviews of Modern Physics}\ }\textbf {\bibinfo {volume} {81}},\
  \bibinfo {pages} {647} (\bibinfo {year} {2009})}\BibitemShut {NoStop}%
\bibitem [{\citenamefont {Lagoudakis}\ \emph {et~al.}(2008)\citenamefont
  {Lagoudakis}, \citenamefont {Wouters}, \citenamefont {Richard}, \citenamefont
  {Baas}, \citenamefont {Carusotto}, \citenamefont {Andr{\'e}}, \citenamefont
  {Dang},\ and\ \citenamefont {Deveaud-Pl{\'e}dran}}]{lagoudakis2008quantized}%
  \BibitemOpen
  \bibfield  {author} {\bibinfo {author} {\bibfnamefont {K.~G.}\ \bibnamefont
  {Lagoudakis}}, \bibinfo {author} {\bibfnamefont {M.}~\bibnamefont {Wouters}},
  \bibinfo {author} {\bibfnamefont {M.}~\bibnamefont {Richard}}, \bibinfo
  {author} {\bibfnamefont {A.}~\bibnamefont {Baas}}, \bibinfo {author}
  {\bibfnamefont {I.}~\bibnamefont {Carusotto}}, \bibinfo {author}
  {\bibfnamefont {R.}~\bibnamefont {Andr{\'e}}}, \bibinfo {author}
  {\bibfnamefont {L.~S.}\ \bibnamefont {Dang}},\ and\ \bibinfo {author}
  {\bibfnamefont {B.}~\bibnamefont {Deveaud-Pl{\'e}dran}},\ }\bibfield  {title}
  {\bibinfo {title} {Quantized vortices in an exciton--polariton condensate},\
  }\href@noop {} {\bibfield  {journal} {\bibinfo  {journal} {Nature physics}\
  }\textbf {\bibinfo {volume} {4}},\ \bibinfo {pages} {706} (\bibinfo {year}
  {2008})}\BibitemShut {NoStop}%
\bibitem [{\citenamefont {Kelvin}(1880)}]{kelvin1880vibrations}%
  \BibitemOpen
  \bibfield  {author} {\bibinfo {author} {\bibfnamefont {L.}~\bibnamefont
  {Kelvin}},\ }\bibfield  {title} {\bibinfo {title} {Vibrations of a columnar
  vortex},\ }\href@noop {} {\bibfield  {journal} {\bibinfo  {journal} {Philos.
  Mag.}\ }\textbf {\bibinfo {volume} {10}},\ \bibinfo {pages} {155} (\bibinfo
  {year} {1880})}\BibitemShut {NoStop}%
\bibitem [{\citenamefont {Pitaevskii}(1961)}]{pitaevskii1961vortex}%
  \BibitemOpen
  \bibfield  {author} {\bibinfo {author} {\bibfnamefont {L.~P.}\ \bibnamefont
  {Pitaevskii}},\ }\bibfield  {title} {\bibinfo {title} {Vortex lines in an
  imperfect bose gas},\ }\href@noop {} {\bibfield  {journal} {\bibinfo
  {journal} {Sov. Phys. JETP}\ }\textbf {\bibinfo {volume} {13}},\ \bibinfo
  {pages} {451} (\bibinfo {year} {1961})}\BibitemShut {NoStop}%
\bibitem [{\citenamefont {Kivotides}\ \emph {et~al.}(2001)\citenamefont
  {Kivotides}, \citenamefont {Vassilicos}, \citenamefont {Samuels},\ and\
  \citenamefont {Barenghi}}]{kivotides2001kelvin}%
  \BibitemOpen
  \bibfield  {author} {\bibinfo {author} {\bibfnamefont {D.}~\bibnamefont
  {Kivotides}}, \bibinfo {author} {\bibfnamefont {J.}~\bibnamefont
  {Vassilicos}}, \bibinfo {author} {\bibfnamefont {D.}~\bibnamefont
  {Samuels}},\ and\ \bibinfo {author} {\bibfnamefont {C.}~\bibnamefont
  {Barenghi}},\ }\bibfield  {title} {\bibinfo {title} {Kelvin waves cascade in
  superfluid turbulence},\ }\href@noop {} {\bibfield  {journal} {\bibinfo
  {journal} {Physical review letters}\ }\textbf {\bibinfo {volume} {86}},\
  \bibinfo {pages} {3080} (\bibinfo {year} {2001})}\BibitemShut {NoStop}%
\bibitem [{\citenamefont {Vinen}\ \emph {et~al.}(2003)\citenamefont {Vinen},
  \citenamefont {Tsubota},\ and\ \citenamefont {Mitani}}]{vinen2003kelvin}%
  \BibitemOpen
  \bibfield  {author} {\bibinfo {author} {\bibfnamefont {W.}~\bibnamefont
  {Vinen}}, \bibinfo {author} {\bibfnamefont {M.}~\bibnamefont {Tsubota}},\
  and\ \bibinfo {author} {\bibfnamefont {A.}~\bibnamefont {Mitani}},\
  }\bibfield  {title} {\bibinfo {title} {Kelvin-wave cascade on a vortex in
  superfluid h e 4 at a very low temperature},\ }\href@noop {} {\bibfield
  {journal} {\bibinfo  {journal} {Physical Review Letters}\ }\textbf {\bibinfo
  {volume} {91}},\ \bibinfo {pages} {135301} (\bibinfo {year}
  {2003})}\BibitemShut {NoStop}%
\bibitem [{\citenamefont {Kozik}\ and\ \citenamefont
  {Svistunov}(2004)}]{kozik2004kelvin}%
  \BibitemOpen
  \bibfield  {author} {\bibinfo {author} {\bibfnamefont {E.}~\bibnamefont
  {Kozik}}\ and\ \bibinfo {author} {\bibfnamefont {B.}~\bibnamefont
  {Svistunov}},\ }\bibfield  {title} {\bibinfo {title} {Kelvin-wave cascade and
  decay of superfluid turbulence},\ }\href@noop {} {\bibfield  {journal}
  {\bibinfo  {journal} {Physical Review Letters}\ }\textbf {\bibinfo {volume}
  {92}},\ \bibinfo {pages} {035301} (\bibinfo {year} {2004})}\BibitemShut
  {NoStop}%
\bibitem [{\citenamefont {L’vov}\ \emph {et~al.}(2006)\citenamefont
  {L’vov}, \citenamefont {Nazarenko},\ and\ \citenamefont
  {Skrbek}}]{l2006energy}%
  \BibitemOpen
  \bibfield  {author} {\bibinfo {author} {\bibfnamefont {V.~S.}\ \bibnamefont
  {L’vov}}, \bibinfo {author} {\bibfnamefont {S.~V.}\ \bibnamefont
  {Nazarenko}},\ and\ \bibinfo {author} {\bibfnamefont {L.}~\bibnamefont
  {Skrbek}},\ }\bibfield  {title} {\bibinfo {title} {Energy spectra of
  developed turbulence in helium superfluids},\ }\href@noop {} {\bibfield
  {journal} {\bibinfo  {journal} {Journal of Low Temperature Physics}\ }\textbf
  {\bibinfo {volume} {145}},\ \bibinfo {pages} {125} (\bibinfo {year}
  {2006})}\BibitemShut {NoStop}%
\bibitem [{\citenamefont {L’vov}\ and\ \citenamefont
  {Nazarenko}(2010)}]{l2010spectrum}%
  \BibitemOpen
  \bibfield  {author} {\bibinfo {author} {\bibfnamefont {V.~S.}\ \bibnamefont
  {L’vov}}\ and\ \bibinfo {author} {\bibfnamefont {S.}~\bibnamefont
  {Nazarenko}},\ }\bibfield  {title} {\bibinfo {title} {Spectrum of kelvin-wave
  turbulence in superfluids},\ }\href@noop {} {\bibfield  {journal} {\bibinfo
  {journal} {JETP Letters}\ }\textbf {\bibinfo {volume} {91}},\ \bibinfo
  {pages} {428} (\bibinfo {year} {2010})}\BibitemShut {NoStop}%
\bibitem [{\citenamefont {Baggaley}\ and\ \citenamefont
  {Barenghi}(2011)}]{baggaley2011spectrum}%
  \BibitemOpen
  \bibfield  {author} {\bibinfo {author} {\bibfnamefont {A.~W.}\ \bibnamefont
  {Baggaley}}\ and\ \bibinfo {author} {\bibfnamefont {C.~F.}\ \bibnamefont
  {Barenghi}},\ }\bibfield  {title} {\bibinfo {title} {Spectrum of turbulent
  kelvin-waves cascade in superfluid helium},\ }\href@noop {} {\bibfield
  {journal} {\bibinfo  {journal} {Physical Review B}\ }\textbf {\bibinfo
  {volume} {83}},\ \bibinfo {pages} {134509} (\bibinfo {year}
  {2011})}\BibitemShut {NoStop}%
\bibitem [{\citenamefont {Sonin}(2012)}]{sonin2012symmetry}%
  \BibitemOpen
  \bibfield  {author} {\bibinfo {author} {\bibfnamefont {E.}~\bibnamefont
  {Sonin}},\ }\bibfield  {title} {\bibinfo {title} {Symmetry of kelvin-wave
  dynamics and the kelvin-wave cascade in the t= 0 superfluid turbulence},\
  }\href@noop {} {\bibfield  {journal} {\bibinfo  {journal} {Physical Review
  B}\ }\textbf {\bibinfo {volume} {85}},\ \bibinfo {pages} {104516} (\bibinfo
  {year} {2012})}\BibitemShut {NoStop}%
\bibitem [{\citenamefont {Autti}\ \emph {et~al.}(2021)\citenamefont {Autti},
  \citenamefont {Heikkinen}, \citenamefont {Laine}, \citenamefont
  {M{\"a}kinen}, \citenamefont {Thuneberg}, \citenamefont {Zavjalov},\ and\
  \citenamefont {Eltsov}}]{autti2021vortex}%
  \BibitemOpen
  \bibfield  {author} {\bibinfo {author} {\bibfnamefont {S.}~\bibnamefont
  {Autti}}, \bibinfo {author} {\bibfnamefont {P.}~\bibnamefont {Heikkinen}},
  \bibinfo {author} {\bibfnamefont {S.}~\bibnamefont {Laine}}, \bibinfo
  {author} {\bibfnamefont {J.}~\bibnamefont {M{\"a}kinen}}, \bibinfo {author}
  {\bibfnamefont {E.}~\bibnamefont {Thuneberg}}, \bibinfo {author}
  {\bibfnamefont {V.}~\bibnamefont {Zavjalov}},\ and\ \bibinfo {author}
  {\bibfnamefont {V.}~\bibnamefont {Eltsov}},\ }\bibfield  {title} {\bibinfo
  {title} {Vortex-mediated relaxation of magnon bec into light higgs
  quasiparticles},\ }\href@noop {} {\bibfield  {journal} {\bibinfo  {journal}
  {Physical Review Research}\ }\textbf {\bibinfo {volume} {3}},\ \bibinfo
  {pages} {L032002} (\bibinfo {year} {2021})}\BibitemShut {NoStop}%
\bibitem [{\citenamefont {Madison}\ \emph {et~al.}(2000)\citenamefont
  {Madison}, \citenamefont {Chevy}, \citenamefont {Wohlleben},\ and\
  \citenamefont {Dalibard}}]{madison2000vortex}%
  \BibitemOpen
  \bibfield  {author} {\bibinfo {author} {\bibfnamefont {K.~W.}\ \bibnamefont
  {Madison}}, \bibinfo {author} {\bibfnamefont {F.}~\bibnamefont {Chevy}},
  \bibinfo {author} {\bibfnamefont {W.}~\bibnamefont {Wohlleben}},\ and\
  \bibinfo {author} {\bibfnamefont {J.}~\bibnamefont {Dalibard}},\ }\bibfield
  {title} {\bibinfo {title} {Vortex formation in a stirred bose-einstein
  condensate},\ }\href@noop {} {\bibfield  {journal} {\bibinfo  {journal}
  {Physical Review Letters}\ }\textbf {\bibinfo {volume} {84}},\ \bibinfo
  {pages} {806} (\bibinfo {year} {2000})}\BibitemShut {NoStop}%
\bibitem [{\citenamefont {Matthews}\ \emph {et~al.}(1999)\citenamefont
  {Matthews}, \citenamefont {Anderson}, \citenamefont {Haljan}, \citenamefont
  {Hall}, \citenamefont {Wieman},\ and\ \citenamefont
  {Cornell}}]{matthews1999vortices}%
  \BibitemOpen
  \bibfield  {author} {\bibinfo {author} {\bibfnamefont {M.~R.}\ \bibnamefont
  {Matthews}}, \bibinfo {author} {\bibfnamefont {B.~P.}\ \bibnamefont
  {Anderson}}, \bibinfo {author} {\bibfnamefont {P.}~\bibnamefont {Haljan}},
  \bibinfo {author} {\bibfnamefont {D.}~\bibnamefont {Hall}}, \bibinfo {author}
  {\bibfnamefont {C.}~\bibnamefont {Wieman}},\ and\ \bibinfo {author}
  {\bibfnamefont {E.~A.}\ \bibnamefont {Cornell}},\ }\bibfield  {title}
  {\bibinfo {title} {Vortices in a bose-einstein condensate},\ }\href@noop {}
  {\bibfield  {journal} {\bibinfo  {journal} {Physical Review Letters}\
  }\textbf {\bibinfo {volume} {83}},\ \bibinfo {pages} {2498} (\bibinfo {year}
  {1999})}\BibitemShut {NoStop}%
\bibitem [{\citenamefont {Neely}\ \emph {et~al.}(2010)\citenamefont {Neely},
  \citenamefont {Samson}, \citenamefont {Bradley}, \citenamefont {Davis},\ and\
  \citenamefont {Anderson}}]{neely2010observation}%
  \BibitemOpen
  \bibfield  {author} {\bibinfo {author} {\bibfnamefont {T.~W.}\ \bibnamefont
  {Neely}}, \bibinfo {author} {\bibfnamefont {E.~C.}\ \bibnamefont {Samson}},
  \bibinfo {author} {\bibfnamefont {A.~S.}\ \bibnamefont {Bradley}}, \bibinfo
  {author} {\bibfnamefont {M.~J.}\ \bibnamefont {Davis}},\ and\ \bibinfo
  {author} {\bibfnamefont {B.~P.}\ \bibnamefont {Anderson}},\ }\bibfield
  {title} {\bibinfo {title} {Observation of vortex dipoles in an oblate
  bose-einstein condensate},\ }\href@noop {} {\bibfield  {journal} {\bibinfo
  {journal} {Physical Review Letters}\ }\textbf {\bibinfo {volume} {104}},\
  \bibinfo {pages} {160401} (\bibinfo {year} {2010})}\BibitemShut {NoStop}%
\bibitem [{\citenamefont {Bretin}\ \emph {et~al.}(2003)\citenamefont {Bretin},
  \citenamefont {Rosenbusch}, \citenamefont {Chevy}, \citenamefont
  {Shlyapnikov},\ and\ \citenamefont {Dalibard}}]{bretin2003quadrupole}%
  \BibitemOpen
  \bibfield  {author} {\bibinfo {author} {\bibfnamefont {V.}~\bibnamefont
  {Bretin}}, \bibinfo {author} {\bibfnamefont {P.}~\bibnamefont {Rosenbusch}},
  \bibinfo {author} {\bibfnamefont {F.}~\bibnamefont {Chevy}}, \bibinfo
  {author} {\bibfnamefont {G.~V.}\ \bibnamefont {Shlyapnikov}},\ and\ \bibinfo
  {author} {\bibfnamefont {J.}~\bibnamefont {Dalibard}},\ }\bibfield  {title}
  {\bibinfo {title} {Quadrupole oscillation of a single-vortex bose-einstein
  condensate: Evidence for kelvin modes},\ }\href@noop {} {\bibfield  {journal}
  {\bibinfo  {journal} {Physical Review Letters}\ }\textbf {\bibinfo {volume}
  {90}},\ \bibinfo {pages} {100403} (\bibinfo {year} {2003})}\BibitemShut
  {NoStop}%
\bibitem [{\citenamefont {Serafini}\ \emph {et~al.}(2015)\citenamefont
  {Serafini}, \citenamefont {Barbiero}, \citenamefont {Debortoli},
  \citenamefont {Donadello}, \citenamefont {Larcher}, \citenamefont {Dalfovo},
  \citenamefont {Lamporesi},\ and\ \citenamefont
  {Ferrari}}]{serafini2015dynamics}%
  \BibitemOpen
  \bibfield  {author} {\bibinfo {author} {\bibfnamefont {S.}~\bibnamefont
  {Serafini}}, \bibinfo {author} {\bibfnamefont {M.}~\bibnamefont {Barbiero}},
  \bibinfo {author} {\bibfnamefont {M.}~\bibnamefont {Debortoli}}, \bibinfo
  {author} {\bibfnamefont {S.}~\bibnamefont {Donadello}}, \bibinfo {author}
  {\bibfnamefont {F.}~\bibnamefont {Larcher}}, \bibinfo {author} {\bibfnamefont
  {F.}~\bibnamefont {Dalfovo}}, \bibinfo {author} {\bibfnamefont
  {G.}~\bibnamefont {Lamporesi}},\ and\ \bibinfo {author} {\bibfnamefont
  {G.}~\bibnamefont {Ferrari}},\ }\bibfield  {title} {\bibinfo {title}
  {Dynamics and interaction of vortex lines in an elongated bose-einstein
  condensate},\ }\href@noop {} {\bibfield  {journal} {\bibinfo  {journal}
  {Physical Review Letters}\ }\textbf {\bibinfo {volume} {115}},\ \bibinfo
  {pages} {170402} (\bibinfo {year} {2015})}\BibitemShut {NoStop}%
\bibitem [{\citenamefont {Serafini}\ \emph {et~al.}(2017)\citenamefont
  {Serafini}, \citenamefont {Galantucci}, \citenamefont {Iseni}, \citenamefont
  {Bienaim{\'e}}, \citenamefont {Bisset}, \citenamefont {Barenghi},
  \citenamefont {Dalfovo}, \citenamefont {Lamporesi},\ and\ \citenamefont
  {Ferrari}}]{serafini2017vortex}%
  \BibitemOpen
  \bibfield  {author} {\bibinfo {author} {\bibfnamefont {S.}~\bibnamefont
  {Serafini}}, \bibinfo {author} {\bibfnamefont {L.}~\bibnamefont
  {Galantucci}}, \bibinfo {author} {\bibfnamefont {E.}~\bibnamefont {Iseni}},
  \bibinfo {author} {\bibfnamefont {T.}~\bibnamefont {Bienaim{\'e}}}, \bibinfo
  {author} {\bibfnamefont {R.~N.}\ \bibnamefont {Bisset}}, \bibinfo {author}
  {\bibfnamefont {C.~F.}\ \bibnamefont {Barenghi}}, \bibinfo {author}
  {\bibfnamefont {F.}~\bibnamefont {Dalfovo}}, \bibinfo {author} {\bibfnamefont
  {G.}~\bibnamefont {Lamporesi}},\ and\ \bibinfo {author} {\bibfnamefont
  {G.}~\bibnamefont {Ferrari}},\ }\bibfield  {title} {\bibinfo {title} {Vortex
  reconnections and rebounds in trapped atomic bose-einstein condensates},\
  }\href@noop {} {\bibfield  {journal} {\bibinfo  {journal} {Physical Review
  X}\ }\textbf {\bibinfo {volume} {7}},\ \bibinfo {pages} {021031} (\bibinfo
  {year} {2017})}\BibitemShut {NoStop}%
\bibitem [{\citenamefont {Fetter}\ and\ \citenamefont
  {Svidzinsky}(2001)}]{fetter2001vortices}%
  \BibitemOpen
  \bibfield  {author} {\bibinfo {author} {\bibfnamefont {A.~L.}\ \bibnamefont
  {Fetter}}\ and\ \bibinfo {author} {\bibfnamefont {A.~A.}\ \bibnamefont
  {Svidzinsky}},\ }\bibfield  {title} {\bibinfo {title} {Vortices in a trapped
  dilute bose-einstein condensate},\ }\href@noop {} {\bibfield  {journal}
  {\bibinfo  {journal} {Journal of Physics: Condensed Matter}\ }\textbf
  {\bibinfo {volume} {13}},\ \bibinfo {pages} {R135} (\bibinfo {year}
  {2001})}\BibitemShut {NoStop}%
\bibitem [{\citenamefont {Fetter}(2004)}]{fetter2004kelvin}%
  \BibitemOpen
  \bibfield  {author} {\bibinfo {author} {\bibfnamefont {A.~L.}\ \bibnamefont
  {Fetter}},\ }\bibfield  {title} {\bibinfo {title} {Kelvin mode of a vortex in
  a nonuniform bose-einstein condensate},\ }\href@noop {} {\bibfield  {journal}
  {\bibinfo  {journal} {Physical Review A}\ }\textbf {\bibinfo {volume} {69}},\
  \bibinfo {pages} {043617} (\bibinfo {year} {2004})}\BibitemShut {NoStop}%
\bibitem [{\citenamefont {Simula}\ \emph
  {et~al.}(2008{\natexlab{a}})\citenamefont {Simula}, \citenamefont
  {Mizushima},\ and\ \citenamefont {Machida}}]{simula2008kelvin}%
  \BibitemOpen
  \bibfield  {author} {\bibinfo {author} {\bibfnamefont {T.~P.}\ \bibnamefont
  {Simula}}, \bibinfo {author} {\bibfnamefont {T.}~\bibnamefont {Mizushima}},\
  and\ \bibinfo {author} {\bibfnamefont {K.}~\bibnamefont {Machida}},\
  }\bibfield  {title} {\bibinfo {title} {Kelvin waves of quantized vortex lines
  in trapped bose-einstein condensates},\ }\href@noop {} {\bibfield  {journal}
  {\bibinfo  {journal} {Physical Review Letters}\ }\textbf {\bibinfo {volume}
  {101}},\ \bibinfo {pages} {020402} (\bibinfo {year}
  {2008}{\natexlab{a}})}\BibitemShut {NoStop}%
\bibitem [{\citenamefont {Simula}\ \emph
  {et~al.}(2008{\natexlab{b}})\citenamefont {Simula}, \citenamefont
  {Mizushima},\ and\ \citenamefont {Machida}}]{simula2008vortex}%
  \BibitemOpen
  \bibfield  {author} {\bibinfo {author} {\bibfnamefont {T.}~\bibnamefont
  {Simula}}, \bibinfo {author} {\bibfnamefont {T.}~\bibnamefont {Mizushima}},\
  and\ \bibinfo {author} {\bibfnamefont {K.}~\bibnamefont {Machida}},\
  }\bibfield  {title} {\bibinfo {title} {Vortex waves in trapped bose-einstein
  condensates},\ }\href@noop {} {\bibfield  {journal} {\bibinfo  {journal}
  {Physical Review A}\ }\textbf {\bibinfo {volume} {78}},\ \bibinfo {pages}
  {053604} (\bibinfo {year} {2008}{\natexlab{b}})}\BibitemShut {NoStop}%
\bibitem [{\citenamefont {Takeuchi}\ \emph {et~al.}(2009)\citenamefont
  {Takeuchi}, \citenamefont {Kasamatsu},\ and\ \citenamefont
  {Tsubota}}]{takeuchi2009spontaneous}%
  \BibitemOpen
  \bibfield  {author} {\bibinfo {author} {\bibfnamefont {H.}~\bibnamefont
  {Takeuchi}}, \bibinfo {author} {\bibfnamefont {K.}~\bibnamefont
  {Kasamatsu}},\ and\ \bibinfo {author} {\bibfnamefont {M.}~\bibnamefont
  {Tsubota}},\ }\bibfield  {title} {\bibinfo {title} {Spontaneous radiation and
  amplification of kelvin waves on quantized vortices in bose-einstein
  condensates},\ }\href@noop {} {\bibfield  {journal} {\bibinfo  {journal}
  {Physical Review A}\ }\textbf {\bibinfo {volume} {79}},\ \bibinfo {pages}
  {033619} (\bibinfo {year} {2009})}\BibitemShut {NoStop}%
\bibitem [{\citenamefont {Simula}\ and\ \citenamefont
  {Machida}(2010)}]{simula2010kelvin}%
  \BibitemOpen
  \bibfield  {author} {\bibinfo {author} {\bibfnamefont {T.}~\bibnamefont
  {Simula}}\ and\ \bibinfo {author} {\bibfnamefont {K.}~\bibnamefont
  {Machida}},\ }\bibfield  {title} {\bibinfo {title} {Kelvin-tkachenko waves of
  few-vortex arrays in trapped bose-einstein condensates},\ }\href@noop {}
  {\bibfield  {journal} {\bibinfo  {journal} {Physical Review A}\ }\textbf
  {\bibinfo {volume} {82}},\ \bibinfo {pages} {063627} (\bibinfo {year}
  {2010})}\BibitemShut {NoStop}%
\bibitem [{\citenamefont {Rooney}\ \emph {et~al.}(2011)\citenamefont {Rooney},
  \citenamefont {Blakie}, \citenamefont {Anderson},\ and\ \citenamefont
  {Bradley}}]{rooney2011suppression}%
  \BibitemOpen
  \bibfield  {author} {\bibinfo {author} {\bibfnamefont {S.}~\bibnamefont
  {Rooney}}, \bibinfo {author} {\bibfnamefont {P.}~\bibnamefont {Blakie}},
  \bibinfo {author} {\bibfnamefont {B.}~\bibnamefont {Anderson}},\ and\
  \bibinfo {author} {\bibfnamefont {A.}~\bibnamefont {Bradley}},\ }\bibfield
  {title} {\bibinfo {title} {Suppression of kelvon-induced decay of quantized
  vortices in oblate bose-einstein condensates},\ }\href@noop {} {\bibfield
  {journal} {\bibinfo  {journal} {Physical Review A}\ }\textbf {\bibinfo
  {volume} {84}},\ \bibinfo {pages} {023637} (\bibinfo {year}
  {2011})}\BibitemShut {NoStop}%
\bibitem [{\citenamefont {Papp}\ \emph {et~al.}(2008)\citenamefont {Papp},
  \citenamefont {Pino},\ and\ \citenamefont {Wieman}}]{papp2008tunable}%
  \BibitemOpen
  \bibfield  {author} {\bibinfo {author} {\bibfnamefont {S.}~\bibnamefont
  {Papp}}, \bibinfo {author} {\bibfnamefont {J.}~\bibnamefont {Pino}},\ and\
  \bibinfo {author} {\bibfnamefont {C.}~\bibnamefont {Wieman}},\ }\bibfield
  {title} {\bibinfo {title} {Tunable miscibility in a dual-species
  bose-einstein condensate},\ }\href@noop {} {\bibfield  {journal} {\bibinfo
  {journal} {Physical Review Letters}\ }\textbf {\bibinfo {volume} {101}},\
  \bibinfo {pages} {040402} (\bibinfo {year} {2008})}\BibitemShut {NoStop}%
\bibitem [{\citenamefont {Tojo}\ \emph {et~al.}(2010)\citenamefont {Tojo},
  \citenamefont {Taguchi}, \citenamefont {Masuyama}, \citenamefont {Hayashi},
  \citenamefont {Saito},\ and\ \citenamefont {Hirano}}]{tojo2010controlling}%
  \BibitemOpen
  \bibfield  {author} {\bibinfo {author} {\bibfnamefont {S.}~\bibnamefont
  {Tojo}}, \bibinfo {author} {\bibfnamefont {Y.}~\bibnamefont {Taguchi}},
  \bibinfo {author} {\bibfnamefont {Y.}~\bibnamefont {Masuyama}}, \bibinfo
  {author} {\bibfnamefont {T.}~\bibnamefont {Hayashi}}, \bibinfo {author}
  {\bibfnamefont {H.}~\bibnamefont {Saito}},\ and\ \bibinfo {author}
  {\bibfnamefont {T.}~\bibnamefont {Hirano}},\ }\bibfield  {title} {\bibinfo
  {title} {Controlling phase separation of binary bose-einstein condensates via
  mixed-spin-channel feshbach resonance},\ }\href@noop {} {\bibfield  {journal}
  {\bibinfo  {journal} {Physical Review A}\ }\textbf {\bibinfo {volume} {82}},\
  \bibinfo {pages} {033609} (\bibinfo {year} {2010})}\BibitemShut {NoStop}%
\bibitem [{\citenamefont {Cabrera}\ \emph {et~al.}(2018)\citenamefont
  {Cabrera}, \citenamefont {Tanzi}, \citenamefont {Sanz}, \citenamefont
  {Naylor}, \citenamefont {Thomas}, \citenamefont {Cheiney},\ and\
  \citenamefont {Tarruell}}]{cabrera2018quantum}%
  \BibitemOpen
  \bibfield  {author} {\bibinfo {author} {\bibfnamefont {C.}~\bibnamefont
  {Cabrera}}, \bibinfo {author} {\bibfnamefont {L.}~\bibnamefont {Tanzi}},
  \bibinfo {author} {\bibfnamefont {J.}~\bibnamefont {Sanz}}, \bibinfo {author}
  {\bibfnamefont {B.}~\bibnamefont {Naylor}}, \bibinfo {author} {\bibfnamefont
  {P.}~\bibnamefont {Thomas}}, \bibinfo {author} {\bibfnamefont
  {P.}~\bibnamefont {Cheiney}},\ and\ \bibinfo {author} {\bibfnamefont
  {L.}~\bibnamefont {Tarruell}},\ }\bibfield  {title} {\bibinfo {title}
  {Quantum liquid droplets in a mixture of bose-einstein condensates},\
  }\href@noop {} {\bibfield  {journal} {\bibinfo  {journal} {Science}\ }\textbf
  {\bibinfo {volume} {359}},\ \bibinfo {pages} {301} (\bibinfo {year}
  {2018})}\BibitemShut {NoStop}%
\bibitem [{\citenamefont {Semeghini}\ \emph {et~al.}(2018)\citenamefont
  {Semeghini}, \citenamefont {Ferioli}, \citenamefont {Masi}, \citenamefont
  {Mazzinghi}, \citenamefont {Wolswijk}, \citenamefont {Minardi}, \citenamefont
  {Modugno}, \citenamefont {Modugno}, \citenamefont {Inguscio},\ and\
  \citenamefont {Fattori}}]{semeghini2018self}%
  \BibitemOpen
  \bibfield  {author} {\bibinfo {author} {\bibfnamefont {G.}~\bibnamefont
  {Semeghini}}, \bibinfo {author} {\bibfnamefont {G.}~\bibnamefont {Ferioli}},
  \bibinfo {author} {\bibfnamefont {L.}~\bibnamefont {Masi}}, \bibinfo {author}
  {\bibfnamefont {C.}~\bibnamefont {Mazzinghi}}, \bibinfo {author}
  {\bibfnamefont {L.}~\bibnamefont {Wolswijk}}, \bibinfo {author}
  {\bibfnamefont {F.}~\bibnamefont {Minardi}}, \bibinfo {author} {\bibfnamefont
  {M.}~\bibnamefont {Modugno}}, \bibinfo {author} {\bibfnamefont
  {G.}~\bibnamefont {Modugno}}, \bibinfo {author} {\bibfnamefont
  {M.}~\bibnamefont {Inguscio}},\ and\ \bibinfo {author} {\bibfnamefont
  {M.}~\bibnamefont {Fattori}},\ }\bibfield  {title} {\bibinfo {title}
  {Self-bound quantum droplets of atomic mixtures in free space},\ }\href@noop
  {} {\bibfield  {journal} {\bibinfo  {journal} {Physical Review Letters}\
  }\textbf {\bibinfo {volume} {120}},\ \bibinfo {pages} {235301} (\bibinfo
  {year} {2018})}\BibitemShut {NoStop}%
\bibitem [{\citenamefont {Skryabin}(2000)}]{skryabin2000instabilities}%
  \BibitemOpen
  \bibfield  {author} {\bibinfo {author} {\bibfnamefont {D.~V.}\ \bibnamefont
  {Skryabin}},\ }\bibfield  {title} {\bibinfo {title} {Instabilities of
  vortices in a binary mixture of trapped bose-einstein condensates: Role of
  collective excitations with positive and negative energies},\ }\href@noop {}
  {\bibfield  {journal} {\bibinfo  {journal} {Physical Review A}\ }\textbf
  {\bibinfo {volume} {63}},\ \bibinfo {pages} {013602} (\bibinfo {year}
  {2000})}\BibitemShut {NoStop}%
\bibitem [{\citenamefont {McGee}\ and\ \citenamefont
  {Holland}(2001)}]{mcgee2001rotational}%
  \BibitemOpen
  \bibfield  {author} {\bibinfo {author} {\bibfnamefont {S.}~\bibnamefont
  {McGee}}\ and\ \bibinfo {author} {\bibfnamefont {M.}~\bibnamefont
  {Holland}},\ }\bibfield  {title} {\bibinfo {title} {Rotational dynamics of
  vortices in confined bose-einstein condensates},\ }\href@noop {} {\bibfield
  {journal} {\bibinfo  {journal} {Physical Review A}\ }\textbf {\bibinfo
  {volume} {63}},\ \bibinfo {pages} {043608} (\bibinfo {year}
  {2001})}\BibitemShut {NoStop}%
\bibitem [{\citenamefont {Eto}\ \emph {et~al.}(2011)\citenamefont {Eto},
  \citenamefont {Kasamatsu}, \citenamefont {Nitta}, \citenamefont {Takeuchi},\
  and\ \citenamefont {Tsubota}}]{eto2011interaction}%
  \BibitemOpen
  \bibfield  {author} {\bibinfo {author} {\bibfnamefont {M.}~\bibnamefont
  {Eto}}, \bibinfo {author} {\bibfnamefont {K.}~\bibnamefont {Kasamatsu}},
  \bibinfo {author} {\bibfnamefont {M.}~\bibnamefont {Nitta}}, \bibinfo
  {author} {\bibfnamefont {H.}~\bibnamefont {Takeuchi}},\ and\ \bibinfo
  {author} {\bibfnamefont {M.}~\bibnamefont {Tsubota}},\ }\bibfield  {title}
  {\bibinfo {title} {Interaction of half-quantized vortices in two-component
  bose-einstein condensates},\ }\href@noop {} {\bibfield  {journal} {\bibinfo
  {journal} {Physical Review A}\ }\textbf {\bibinfo {volume} {83}},\ \bibinfo
  {pages} {063603} (\bibinfo {year} {2011})}\BibitemShut {NoStop}%
\bibitem [{\citenamefont {Aioi}\ \emph {et~al.}(2012)\citenamefont {Aioi},
  \citenamefont {Kadokura},\ and\ \citenamefont {Saito}}]{aioi2012penetration}%
  \BibitemOpen
  \bibfield  {author} {\bibinfo {author} {\bibfnamefont {T.}~\bibnamefont
  {Aioi}}, \bibinfo {author} {\bibfnamefont {T.}~\bibnamefont {Kadokura}},\
  and\ \bibinfo {author} {\bibfnamefont {H.}~\bibnamefont {Saito}},\ }\bibfield
   {title} {\bibinfo {title} {Penetration of a vortex dipole across an
  interface of bose-einstein condensates},\ }\href@noop {} {\bibfield
  {journal} {\bibinfo  {journal} {Physical Review A}\ }\textbf {\bibinfo
  {volume} {85}},\ \bibinfo {pages} {023618} (\bibinfo {year}
  {2012})}\BibitemShut {NoStop}%
\bibitem [{\citenamefont {Ishino}\ \emph {et~al.}(2013)\citenamefont {Ishino},
  \citenamefont {Tsubota},\ and\ \citenamefont {Takeuchi}}]{ishino2013counter}%
  \BibitemOpen
  \bibfield  {author} {\bibinfo {author} {\bibfnamefont {S.}~\bibnamefont
  {Ishino}}, \bibinfo {author} {\bibfnamefont {M.}~\bibnamefont {Tsubota}},\
  and\ \bibinfo {author} {\bibfnamefont {H.}~\bibnamefont {Takeuchi}},\
  }\bibfield  {title} {\bibinfo {title} {Counter-rotating vortices in miscible
  two-component bose-einstein condensates},\ }\href@noop {} {\bibfield
  {journal} {\bibinfo  {journal} {Physical Review A}\ }\textbf {\bibinfo
  {volume} {88}},\ \bibinfo {pages} {063617} (\bibinfo {year}
  {2013})}\BibitemShut {NoStop}%
\bibitem [{\citenamefont {Kasamatsu}\ \emph {et~al.}(2016)\citenamefont
  {Kasamatsu}, \citenamefont {Eto},\ and\ \citenamefont
  {Nitta}}]{kasamatsu2016short}%
  \BibitemOpen
  \bibfield  {author} {\bibinfo {author} {\bibfnamefont {K.}~\bibnamefont
  {Kasamatsu}}, \bibinfo {author} {\bibfnamefont {M.}~\bibnamefont {Eto}},\
  and\ \bibinfo {author} {\bibfnamefont {M.}~\bibnamefont {Nitta}},\ }\bibfield
   {title} {\bibinfo {title} {Short-range intervortex interaction and
  interacting dynamics of half-quantized vortices in two-component
  bose-einstein condensates},\ }\href@noop {} {\bibfield  {journal} {\bibinfo
  {journal} {Physical Review A}\ }\textbf {\bibinfo {volume} {93}},\ \bibinfo
  {pages} {013615} (\bibinfo {year} {2016})}\BibitemShut {NoStop}%
\bibitem [{\citenamefont {Wang}\ \emph {et~al.}(2018)\citenamefont {Wang},
  \citenamefont {Dai}, \citenamefont {Wen}, \citenamefont {Liu}, \citenamefont
  {Jiang}, \citenamefont {Saito}, \citenamefont {Zhang},\ and\ \citenamefont
  {Zhang}}]{wang2018dynamics}%
  \BibitemOpen
  \bibfield  {author} {\bibinfo {author} {\bibfnamefont {L.-X.}\ \bibnamefont
  {Wang}}, \bibinfo {author} {\bibfnamefont {C.-Q.}\ \bibnamefont {Dai}},
  \bibinfo {author} {\bibfnamefont {L.}~\bibnamefont {Wen}}, \bibinfo {author}
  {\bibfnamefont {T.}~\bibnamefont {Liu}}, \bibinfo {author} {\bibfnamefont
  {H.-F.}\ \bibnamefont {Jiang}}, \bibinfo {author} {\bibfnamefont
  {H.}~\bibnamefont {Saito}}, \bibinfo {author} {\bibfnamefont {S.-G.}\
  \bibnamefont {Zhang}},\ and\ \bibinfo {author} {\bibfnamefont {X.-F.}\
  \bibnamefont {Zhang}},\ }\bibfield  {title} {\bibinfo {title} {Dynamics of
  vortices followed by the collapse of ring dark solitons in a two-component
  bose-einstein condensate},\ }\href@noop {} {\bibfield  {journal} {\bibinfo
  {journal} {Physical Review A}\ }\textbf {\bibinfo {volume} {97}},\ \bibinfo
  {pages} {063607} (\bibinfo {year} {2018})}\BibitemShut {NoStop}%
\bibitem [{\citenamefont {Kuopanportti}\ \emph {et~al.}(2019)\citenamefont
  {Kuopanportti}, \citenamefont {Bandyopadhyay}, \citenamefont {Roy},\ and\
  \citenamefont {Angom}}]{kuopanportti2019splitting}%
  \BibitemOpen
  \bibfield  {author} {\bibinfo {author} {\bibfnamefont {P.}~\bibnamefont
  {Kuopanportti}}, \bibinfo {author} {\bibfnamefont {S.}~\bibnamefont
  {Bandyopadhyay}}, \bibinfo {author} {\bibfnamefont {A.}~\bibnamefont {Roy}},\
  and\ \bibinfo {author} {\bibfnamefont {D.}~\bibnamefont {Angom}},\ }\bibfield
   {title} {\bibinfo {title} {Splitting of singly and doubly quantized
  composite vortices in two-component bose-einstein condensates},\ }\href@noop
  {} {\bibfield  {journal} {\bibinfo  {journal} {Physical Review A}\ }\textbf
  {\bibinfo {volume} {100}},\ \bibinfo {pages} {033615} (\bibinfo {year}
  {2019})}\BibitemShut {NoStop}%
\bibitem [{\citenamefont {Richaud}\ \emph {et~al.}(2021)\citenamefont
  {Richaud}, \citenamefont {Penna},\ and\ \citenamefont
  {Fetter}}]{richaud2021dynamics}%
  \BibitemOpen
  \bibfield  {author} {\bibinfo {author} {\bibfnamefont {A.}~\bibnamefont
  {Richaud}}, \bibinfo {author} {\bibfnamefont {V.}~\bibnamefont {Penna}},\
  and\ \bibinfo {author} {\bibfnamefont {A.~L.}\ \bibnamefont {Fetter}},\
  }\bibfield  {title} {\bibinfo {title} {Dynamics of massive point vortices in
  a binary mixture of bose-einstein condensates},\ }\href@noop {} {\bibfield
  {journal} {\bibinfo  {journal} {Physical Review A}\ }\textbf {\bibinfo
  {volume} {103}},\ \bibinfo {pages} {023311} (\bibinfo {year}
  {2021})}\BibitemShut {NoStop}%
\bibitem [{\citenamefont {Han}\ and\ \citenamefont
  {Tsubota}(2021)}]{han2021annihilation}%
  \BibitemOpen
  \bibfield  {author} {\bibinfo {author} {\bibfnamefont {J.}~\bibnamefont
  {Han}}\ and\ \bibinfo {author} {\bibfnamefont {M.}~\bibnamefont {Tsubota}},\
  }\bibfield  {title} {\bibinfo {title} {Annihilation and recurrence of
  vortex-antivortex pairs in two-component bose-einstein condensates},\
  }\href@noop {} {\bibfield  {journal} {\bibinfo  {journal} {Physical Review
  A}\ }\textbf {\bibinfo {volume} {103}},\ \bibinfo {pages} {053313} (\bibinfo
  {year} {2021})}\BibitemShut {NoStop}%
\bibitem [{\citenamefont {Edmonds}\ \emph {et~al.}(2021)\citenamefont
  {Edmonds}, \citenamefont {Eto},\ and\ \citenamefont
  {Nitta}}]{edmonds2021synthetic}%
  \BibitemOpen
  \bibfield  {author} {\bibinfo {author} {\bibfnamefont {M.}~\bibnamefont
  {Edmonds}}, \bibinfo {author} {\bibfnamefont {M.}~\bibnamefont {Eto}},\ and\
  \bibinfo {author} {\bibfnamefont {M.}~\bibnamefont {Nitta}},\ }\bibfield
  {title} {\bibinfo {title} {Synthetic superfluid chemistry with vortex-trapped
  quantum impurities},\ }\href@noop {} {\bibfield  {journal} {\bibinfo
  {journal} {Physical Review Research}\ }\textbf {\bibinfo {volume} {3}},\
  \bibinfo {pages} {023085} (\bibinfo {year} {2021})}\BibitemShut {NoStop}%
\bibitem [{\citenamefont {Ruban}(2021{\natexlab{a}})}]{ruban2021instabilities}%
  \BibitemOpen
  \bibfield  {author} {\bibinfo {author} {\bibfnamefont {V.~P.}\ \bibnamefont
  {Ruban}},\ }\bibfield  {title} {\bibinfo {title} {Instabilities of a filled
  vortex in a two-component bose--einstein condensate},\ }\href@noop {}
  {\bibfield  {journal} {\bibinfo  {journal} {JETP Letters}\ }\textbf {\bibinfo
  {volume} {113}},\ \bibinfo {pages} {532} (\bibinfo {year}
  {2021}{\natexlab{a}})}\BibitemShut {NoStop}%
\bibitem [{\citenamefont {Ruban}(2021{\natexlab{b}})}]{ruban2021bubbles}%
  \BibitemOpen
  \bibfield  {author} {\bibinfo {author} {\bibfnamefont {V.~P.}\ \bibnamefont
  {Ruban}},\ }\bibfield  {title} {\bibinfo {title} {Bubbles with attached
  quantum vortices in trapped binary bose--einstein condensates},\ }\href@noop
  {} {\bibfield  {journal} {\bibinfo  {journal} {Journal of Experimental and
  Theoretical Physics}\ }\textbf {\bibinfo {volume} {133}},\ \bibinfo {pages}
  {779} (\bibinfo {year} {2021}{\natexlab{b}})}\BibitemShut {NoStop}%
\bibitem [{\citenamefont {Han}\ \emph {et~al.}(2022)\citenamefont {Han},
  \citenamefont {Kasamatsu},\ and\ \citenamefont {Tsubota}}]{han2022dynamics}%
  \BibitemOpen
  \bibfield  {author} {\bibinfo {author} {\bibfnamefont {J.}~\bibnamefont
  {Han}}, \bibinfo {author} {\bibfnamefont {K.}~\bibnamefont {Kasamatsu}},\
  and\ \bibinfo {author} {\bibfnamefont {M.}~\bibnamefont {Tsubota}},\
  }\bibfield  {title} {\bibinfo {title} {Dynamics of two quantized vortices
  belonging to different components of binary bose--einstein condensates in a
  circular box potential},\ }\href@noop {} {\bibfield  {journal} {\bibinfo
  {journal} {Journal of the Physical Society of Japan}\ }\textbf {\bibinfo
  {volume} {91}},\ \bibinfo {pages} {024401} (\bibinfo {year}
  {2022})}\BibitemShut {NoStop}%
\bibitem [{\citenamefont {Hayashi}\ \emph {et~al.}(2013)\citenamefont
  {Hayashi}, \citenamefont {Tsubota},\ and\ \citenamefont
  {Takeuchi}}]{hayashi2013instability}%
  \BibitemOpen
  \bibfield  {author} {\bibinfo {author} {\bibfnamefont {S.}~\bibnamefont
  {Hayashi}}, \bibinfo {author} {\bibfnamefont {M.}~\bibnamefont {Tsubota}},\
  and\ \bibinfo {author} {\bibfnamefont {H.}~\bibnamefont {Takeuchi}},\
  }\bibfield  {title} {\bibinfo {title} {Instability crossover of helical shear
  flow in segregated bose-einstein condensates},\ }\href@noop {} {\bibfield
  {journal} {\bibinfo  {journal} {Physical Review A}\ }\textbf {\bibinfo
  {volume} {87}},\ \bibinfo {pages} {063628} (\bibinfo {year}
  {2013})}\BibitemShut {NoStop}%
\bibitem [{\citenamefont {Kasamatsu}\ \emph {et~al.}(2005)\citenamefont
  {Kasamatsu}, \citenamefont {Tsubota},\ and\ \citenamefont
  {Ueda}}]{kasamatsu2005vortices}%
  \BibitemOpen
  \bibfield  {author} {\bibinfo {author} {\bibfnamefont {K.}~\bibnamefont
  {Kasamatsu}}, \bibinfo {author} {\bibfnamefont {M.}~\bibnamefont {Tsubota}},\
  and\ \bibinfo {author} {\bibfnamefont {M.}~\bibnamefont {Ueda}},\ }\bibfield
  {title} {\bibinfo {title} {Vortices in multicomponent bose--einstein
  condensates},\ }\href@noop {} {\bibfield  {journal} {\bibinfo  {journal}
  {International Journal of Modern Physics B}\ }\textbf {\bibinfo {volume}
  {19}},\ \bibinfo {pages} {1835} (\bibinfo {year} {2005})}\BibitemShut
  {NoStop}%
\bibitem [{\citenamefont {Kuopanportti}\ \emph {et~al.}(2012)\citenamefont
  {Kuopanportti}, \citenamefont {Huhtam{\"a}ki},\ and\ \citenamefont
  {M{\"o}tt{\"o}nen}}]{kuopanportti2012exotic}%
  \BibitemOpen
  \bibfield  {author} {\bibinfo {author} {\bibfnamefont {P.}~\bibnamefont
  {Kuopanportti}}, \bibinfo {author} {\bibfnamefont {J.~A.}\ \bibnamefont
  {Huhtam{\"a}ki}},\ and\ \bibinfo {author} {\bibfnamefont {M.}~\bibnamefont
  {M{\"o}tt{\"o}nen}},\ }\bibfield  {title} {\bibinfo {title} {Exotic vortex
  lattices in two-species bose-einstein condensates},\ }\href@noop {}
  {\bibfield  {journal} {\bibinfo  {journal} {Physical Review A}\ }\textbf
  {\bibinfo {volume} {85}},\ \bibinfo {pages} {043613} (\bibinfo {year}
  {2012})}\BibitemShut {NoStop}%
\bibitem [{\citenamefont {Pu}\ \emph {et~al.}(1999)\citenamefont {Pu},
  \citenamefont {Law}, \citenamefont {Eberly},\ and\ \citenamefont
  {Bigelow}}]{pu1999coherent}%
  \BibitemOpen
  \bibfield  {author} {\bibinfo {author} {\bibfnamefont {H.}~\bibnamefont
  {Pu}}, \bibinfo {author} {\bibfnamefont {C.}~\bibnamefont {Law}}, \bibinfo
  {author} {\bibfnamefont {J.}~\bibnamefont {Eberly}},\ and\ \bibinfo {author}
  {\bibfnamefont {N.}~\bibnamefont {Bigelow}},\ }\bibfield  {title} {\bibinfo
  {title} {Coherent disintegration and stability of vortices in trapped bose
  condensates},\ }\href@noop {} {\bibfield  {journal} {\bibinfo  {journal}
  {Physical Review A}\ }\textbf {\bibinfo {volume} {59}},\ \bibinfo {pages}
  {1533} (\bibinfo {year} {1999})}\BibitemShut {NoStop}%
\bibitem [{\citenamefont {M{\"o}tt{\"o}nen}\ \emph {et~al.}(2003)\citenamefont
  {M{\"o}tt{\"o}nen}, \citenamefont {Mizushima}, \citenamefont {Isoshima},
  \citenamefont {Salomaa},\ and\ \citenamefont
  {Machida}}]{mottonen2003splitting}%
  \BibitemOpen
  \bibfield  {author} {\bibinfo {author} {\bibfnamefont {M.}~\bibnamefont
  {M{\"o}tt{\"o}nen}}, \bibinfo {author} {\bibfnamefont {T.}~\bibnamefont
  {Mizushima}}, \bibinfo {author} {\bibfnamefont {T.}~\bibnamefont {Isoshima}},
  \bibinfo {author} {\bibfnamefont {M.~M.}\ \bibnamefont {Salomaa}},\ and\
  \bibinfo {author} {\bibfnamefont {K.}~\bibnamefont {Machida}},\ }\bibfield
  {title} {\bibinfo {title} {Splitting of a doubly quantized vortex through
  intertwining in bose-einstein condensates},\ }\href@noop {} {\bibfield
  {journal} {\bibinfo  {journal} {Physical Review A}\ }\textbf {\bibinfo
  {volume} {68}},\ \bibinfo {pages} {023611} (\bibinfo {year}
  {2003})}\BibitemShut {NoStop}%
\bibitem [{\citenamefont {Kawaguchi}\ and\ \citenamefont
  {Ohmi}(2004)}]{kawaguchi2004splitting}%
  \BibitemOpen
  \bibfield  {author} {\bibinfo {author} {\bibfnamefont {Y.}~\bibnamefont
  {Kawaguchi}}\ and\ \bibinfo {author} {\bibfnamefont {T.}~\bibnamefont
  {Ohmi}},\ }\bibfield  {title} {\bibinfo {title} {Splitting instability of a
  multiply charged vortex in a bose-einstein condensate},\ }\href@noop {}
  {\bibfield  {journal} {\bibinfo  {journal} {Physical Review A}\ }\textbf
  {\bibinfo {volume} {70}},\ \bibinfo {pages} {043610} (\bibinfo {year}
  {2004})}\BibitemShut {NoStop}%
\bibitem [{\citenamefont {Shin}\ \emph {et~al.}(2004)\citenamefont {Shin},
  \citenamefont {Saba}, \citenamefont {Vengalattore}, \citenamefont {Pasquini},
  \citenamefont {Sanner}, \citenamefont {Leanhardt}, \citenamefont {Prentiss},
  \citenamefont {Pritchard},\ and\ \citenamefont
  {Ketterle}}]{shin2004dynamical}%
  \BibitemOpen
  \bibfield  {author} {\bibinfo {author} {\bibfnamefont {Y.-i.}\ \bibnamefont
  {Shin}}, \bibinfo {author} {\bibfnamefont {M.}~\bibnamefont {Saba}}, \bibinfo
  {author} {\bibfnamefont {M.}~\bibnamefont {Vengalattore}}, \bibinfo {author}
  {\bibfnamefont {T.}~\bibnamefont {Pasquini}}, \bibinfo {author}
  {\bibfnamefont {C.}~\bibnamefont {Sanner}}, \bibinfo {author} {\bibfnamefont
  {A.}~\bibnamefont {Leanhardt}}, \bibinfo {author} {\bibfnamefont
  {M.}~\bibnamefont {Prentiss}}, \bibinfo {author} {\bibfnamefont
  {D.}~\bibnamefont {Pritchard}},\ and\ \bibinfo {author} {\bibfnamefont
  {W.}~\bibnamefont {Ketterle}},\ }\bibfield  {title} {\bibinfo {title}
  {Dynamical instability of a doubly quantized vortex in a bose-einstein
  condensate},\ }\href@noop {} {\bibfield  {journal} {\bibinfo  {journal}
  {Physical Review Letters}\ }\textbf {\bibinfo {volume} {93}},\ \bibinfo
  {pages} {160406} (\bibinfo {year} {2004})}\BibitemShut {NoStop}%
\bibitem [{\citenamefont {Lundh}\ and\ \citenamefont
  {Nilsen}(2006)}]{lundh2006dynamic}%
  \BibitemOpen
  \bibfield  {author} {\bibinfo {author} {\bibfnamefont {E.}~\bibnamefont
  {Lundh}}\ and\ \bibinfo {author} {\bibfnamefont {H.~M.}\ \bibnamefont
  {Nilsen}},\ }\bibfield  {title} {\bibinfo {title} {Dynamic stability of a
  doubly quantized vortex in a three-dimensional condensate},\ }\href@noop {}
  {\bibfield  {journal} {\bibinfo  {journal} {Physical Review A}\ }\textbf
  {\bibinfo {volume} {74}},\ \bibinfo {pages} {063620} (\bibinfo {year}
  {2006})}\BibitemShut {NoStop}%
\bibitem [{\citenamefont {Takeuchi}\ \emph {et~al.}(2018)\citenamefont
  {Takeuchi}, \citenamefont {Kobayashi},\ and\ \citenamefont
  {Kasamatsu}}]{takeuchi2018doubly}%
  \BibitemOpen
  \bibfield  {author} {\bibinfo {author} {\bibfnamefont {H.}~\bibnamefont
  {Takeuchi}}, \bibinfo {author} {\bibfnamefont {M.}~\bibnamefont
  {Kobayashi}},\ and\ \bibinfo {author} {\bibfnamefont {K.}~\bibnamefont
  {Kasamatsu}},\ }\bibfield  {title} {\bibinfo {title} {Is a doubly quantized
  vortex dynamically unstable in uniform superfluids?},\ }\href@noop {}
  {\bibfield  {journal} {\bibinfo  {journal} {Journal of the Physical Society
  of Japan}\ }\textbf {\bibinfo {volume} {87}},\ \bibinfo {pages} {023601}
  (\bibinfo {year} {2018})}\BibitemShut {NoStop}%
\bibitem [{\citenamefont {Kobayashi}\ and\ \citenamefont
  {Nitta}(2014)}]{kobayashi2014kelvin}%
  \BibitemOpen
  \bibfield  {author} {\bibinfo {author} {\bibfnamefont {M.}~\bibnamefont
  {Kobayashi}}\ and\ \bibinfo {author} {\bibfnamefont {M.}~\bibnamefont
  {Nitta}},\ }\bibfield  {title} {\bibinfo {title} {Kelvin modes as
  nambu--goldstone modes along superfluid vortices and relativistic strings:
  Finite volume size effects},\ }\href@noop {} {\bibfield  {journal} {\bibinfo
  {journal} {Progress of Theoretical and Experimental Physics}\ }\textbf
  {\bibinfo {volume} {2014}},\ \bibinfo {pages} {021B01} (\bibinfo {year}
  {2014})}\BibitemShut {NoStop}%
\bibitem [{\citenamefont {Takahashi}\ \emph {et~al.}(2015)\citenamefont
  {Takahashi}, \citenamefont {Kobayashi},\ and\ \citenamefont
  {Nitta}}]{takahashi2015nambu}%
  \BibitemOpen
  \bibfield  {author} {\bibinfo {author} {\bibfnamefont {D.~A.}\ \bibnamefont
  {Takahashi}}, \bibinfo {author} {\bibfnamefont {M.}~\bibnamefont
  {Kobayashi}},\ and\ \bibinfo {author} {\bibfnamefont {M.}~\bibnamefont
  {Nitta}},\ }\bibfield  {title} {\bibinfo {title} {Nambu-goldstone modes
  propagating along topological defects: Kelvin and ripple modes from small to
  large systems},\ }\href@noop {} {\bibfield  {journal} {\bibinfo  {journal}
  {Physical Review B}\ }\textbf {\bibinfo {volume} {91}},\ \bibinfo {pages}
  {184501} (\bibinfo {year} {2015})}\BibitemShut {NoStop}%
\bibitem [{\citenamefont {Ao}\ and\ \citenamefont {Chui}(1998)}]{ao1998binary}%
  \BibitemOpen
  \bibfield  {author} {\bibinfo {author} {\bibfnamefont {P.}~\bibnamefont
  {Ao}}\ and\ \bibinfo {author} {\bibfnamefont {S.}~\bibnamefont {Chui}},\
  }\bibfield  {title} {\bibinfo {title} {Binary bose-einstein condensate
  mixtures in weakly and strongly segregated phases},\ }\href@noop {}
  {\bibfield  {journal} {\bibinfo  {journal} {Physical Review A}\ }\textbf
  {\bibinfo {volume} {58}},\ \bibinfo {pages} {4836} (\bibinfo {year}
  {1998})}\BibitemShut {NoStop}%
\bibitem [{\citenamefont {Pethick}\ and\ \citenamefont
  {Smith}(2008)}]{pethick2008bose}%
  \BibitemOpen
  \bibfield  {author} {\bibinfo {author} {\bibfnamefont {C.~J.}\ \bibnamefont
  {Pethick}}\ and\ \bibinfo {author} {\bibfnamefont {H.}~\bibnamefont
  {Smith}},\ }\href@noop {} {\emph {\bibinfo {title} {Bose--Einstein
  condensation in dilute gases}}}\ (\bibinfo  {publisher} {Cambridge university
  press},\ \bibinfo {year} {2008})\BibitemShut {NoStop}%
\end{thebibliography}%

\end{document}